\documentclass[onecolumn,unpublished,letterpaper]{quantumarticle}

\usepackage{tikz}
\usepackage[sort, numbers]{natbib}
\usepackage{revquantum}

\newcommand{\figurefolder}{../fig}

\newcommand{\variance}{\mathrm{Var}}

\newcommand{\ee}{\mathrm{e}}

\newcommand{\lhood}{\mathrm{L}}

\newcommand{\binomial}{\operatorname{Binom}}
\newcommand{\uniform}{\operatorname{Unif}}
\newcommand{\normal}{\operatorname{N}}
\newcommand{\betabin}{\operatorname{Beta-Binom}}
\newcommand{\betafun}[1]{\operatorname{B}{\left({#1}\right)}}
\newcommand{\betadist}{\operatorname{Beta}}
\newcommand{\dirichlet}{\operatorname{Dir}}

\newcommand{\poisson}{\operatorname{Pois}}
\newcommand{\gammadist}{\operatorname{Gam}}
\newcommand{\dirp}{\operatorname{DP}}
\newcommand{\CDPBM}{\operatorname{CDPBM}}

\newcommand\iid{~\stackrel{\mathclap{\normalfont\mbox{\small{iid}}}}{\sim}~}
\newcommand\ind{~\stackrel{\mathclap{\normalfont\mbox{\small{ind}}}}{\sim}~}

\newcommand{\Gset}{\mathbb{G}}
\newcommand{\G}{\mathcal{G}}
\newcommand{\M}{\mathbb{M}}
\newcommand{\Eset}{\mathfrak{E}}
\newcommand{\E}{\mathcal{E}}
\newcommand{\X}{\mathcal{X}}

\newcommand{\err}{{\text{err}}}

\newcommand{\PAL}{{\text{PAL}}}

\renewcommand{\figurefolder}{./fig}
\begin{document}


\title{Bayesian Inference for \\Randomized Benchmarking Protocols}

\author{Ian Hincks}
\email{ihincks@uwaterloo.ca}
\affilUWAMath
\affilIQC

\author{Joel J. Wallman}
\affilUWAMath
\affilIQC

\author{Chris Ferrie}
\affilUTechSyd

\author{Chris Granade}
\affilQuArC

\author{David G. Cory}
\affilUWChem
\affilIQC
\affilPI
\affilCIFAR

\date{\today}



\begin{abstract}
Randomized benchmarking (RB) protocols are standard tools for characterizing quantum devices.
Prior analyses of RB protocols have not provided a complete method for analyzing realistic data, resulting in a variety of ad-hoc methods.
The main confounding factor in rigorously analyzing data from RB protocols is an unknown and noise-dependent distribution of survival probabilities over random sequences.
We propose a hierarchical Bayesian method where these survival distributions
are modeled as nonparametric Dirichlet process mixtures.
Our method infers parameters of interest without additional assumptions about the underlying physical noise process.
We show with numerical examples that our method works robustly for both standard and highly pathological error models.
Our method also works reliably at low noise levels and with little data because 
we avoid the asymptotic assumptions of commonly used methods such as 
least-squares fitting. For example, our method produces a narrow and consistent
posterior for the average gate fidelity from ten random sequences per sequence length in the standard RB protocol.
\end{abstract}


\maketitle

\section{Introduction}

Accurately characterizing the performance of both large and small quantum
devices is vital to ensure that, for example, quantum information processors are 
reliable and metrology devices are accurate.
For critical applications, the reliability of confidence intervals or credible regions for figures of merit is more important than a single-point estimate as there might be practical consequences to over-reporting the performance of a device.

Currently, the only known scalable protocols for characterizing discrete quantum logic gates are randomized benchmarking (RB)~\cite{
emerson_scalable_2005,
knill_randomized_2008,
magesan_scalable_2011,
magesan_characterizing_2012} and variants thereof, collectively referred to as RB+ (see \autoref{tab:models} for some variants).
The standard RB protocol works by applying random sequences of gates that
ideally compose to the identity, where the gates form a unitary
2-design~\cite{dankert_exact_2009}.
Measuring in the basis of any initial state after applying a random sequence
therefore gives an estimate of the survival probability conditioned upon that
random sequence.
The survival probability averaged over all random sequences of a fixed length
decays exponentially with the length, where the decay rate is a linear function
of the average gate fidelity of the overall noise channel.
Members of RB+ all have similar structure, modified to suit different goals.
RB has been experimentally implemented on a large variety of quantum
platforms~\cite{
brown_single-qubit-gate_2011,
moussa_practical_2012,
veldhorst_addressable_2014,
barends_superconducting_2014,
muhonen_quantifying_2015,
xia_randomized_2015,
sheldon_characterizing_2016,
feng_estimating_2016,
hro+_implementing_2016}
, and is so ubiquitous that its results are often
reported with little detail within the context of a larger purpose.

However, these experimental implementations make different ad-hoc statistical
assumptions because previous theoretical treatments of RB+ have typically
neglected data analysis.
The analysis of RB+ experiments is complicated by three factors:
\begin{enumerate}
    \item every random sequence in a protocol gives rise to a different survival
    probability, giving rise to a survival distribution for each sequence
    length;
    \item in low- to mid-data regimes, assuming Gaussian errors on either
    the estimates of the individual survival probabilities or on the mean
    of the survival distribution through the central limit theorem is dubious;
    and
    \item applying hard physical constraints violates the assumptions of
    standard statistical fitting routines.
\end{enumerate}

This paper presents a Bayesian data-processing method that
overcomes these difficulties, and that can be applied to 
all members of RB+.
As with any Bayesian approach, the output is a joint posterior distribution
over all parameters relevant to the problem.
Joint distributions over the parameter(s) of interest can be obtained by
marginalizing over nuisance parameters, enabling straight-forward statements
like `under this protocol's model with this prior knowledge, there is a 95\%
probability that such-and-such parameter is greater than 0.999'.
If a point estimate is required for some parameter, the Bayes estimate is just a
sum and division away.

This paper is organized as follows.
In \autoref{sec:framework} we layout a notational framework for RB+.
In \autoref{sec:survival-distributions} we discuss how every protocol relates 
survival distributions of different lengths through what we call 
\textit{tying functions}.
This leads to the likelihood function of RB+ defined in
\autoref{sec:likelihood-function}, and its necessary dependence on moments of 
the survival distributions.
This is used in \autoref{sec:constructing-agnostic-models} to define and 
motivate our main Bayesian model, along with alternative frequentist approaches.
We then discuss optimal sequence re-use strategies in
\autoref{sec:sequence-re-use}.
We present the results of numerical simulations in
\autoref{sec:numerical-results}.
Finally, in \autoref{sec:departing-from-bernoulli-trials}, we briefly discuss
how our model can be extended to systems or protocols without strong two-outcome
measurements.

\section{The Framework of RB+}
\label{sec:framework}

In this section we provide a general framework to rapidly understand and compare 
the various protocols related to randomized benchmarking. 
The framework consists of the following six elements, exemplified in
\autoref{tab:models}:
\begin{enumerate}
    \item $\Gset$, Gate Set: the set of $R$ gates $\Gset=\{G_1,G_2,...,G_R\}$
    used in the protocol, where this set might satisfy specific conditions such
    as being a group and a unitary 2-design\footnote{As a point of 
    practicality, note that gates from $\Gset$ are often physically
    implemented by compiling gates from a smaller generating set of 
    gates that need not share any special properties required 
    by $\Gset$.};
    \item $\Eset$, Experiment Types: labels for protocols that combine data from
    multiple sub-protocols, possibly including specification of multiple
    configurations of preparation and measurement (SPAM), denoted
    with $\rho$ and $E$ respectively\footnote{Rather
    than including SPAM configurations as experiment types, sometimes
    protocols may instead compile SPAM configurations into the
    allowable sequences.};
    \item $M$, the Sequence Length: a positive integer, where $\ell(M,e)$ 
    denotes the exact number
    of gates from $\Gset$ needed to construct a sequence at length 
    $M$ under experiment type $e\in\Eset$;
    \item $ J_{M,e}$, Allowable Sequences: a discrete distribution 
    whose sample space is the set of gate-indexing tuples 
    $\{1,...,R\}^{\ell(M,e)}$, typically uniform on a subset thereof;
    \item $ \vec{x}_T$, Tying Parameters: the set of parameters that can be learned from the protocol; and,
    \item $T$, Tying Functions: the known dependence of the parameters on the statistics of the measurement data.
\end{enumerate}

For a given sequence of gate indices
$\vec{j}=(j_1,...,j_{K})$, define the corresponding ideal gate as
\begin{equation}
    \G_{\vec{j}}=\G_{j_K} \cdots \G_{j_2}\G_{j_1},
\end{equation}
where we use the convention that the scripted version of a letter denoting a 
unitary operator is the quantum channel which conjugates by that unitary,
that is, $\G(\rho)=G\rho G^\dagger$.
We write the imperfect implementations of $\rho$, $\G_{\vec{j}}$, and $E$ as
$\tilde{\rho}$, $\tilde{\G}_{\vec{j}}$, and $\tilde{E}$ respectively.
The following procedure is then performed experimentally, possibly in a random
order to prevent experimental drifts from causing a systematic error:

\algrenewcommand\algorithmicindent{1em}
\begin{algorithmic}
\For {each sequence length $M\in\M$}
    \For {each experiment type $e\in\Eset_M}$
        \For {each sequence $i=1,2,3,...,I$}
          \State $\vec{j} \gets \RV{J_{M,e}}$
            \State $Q_{M,e,i} \gets \RV{\Binom{N}{
                \Tr \tilde{E}_e \tilde{\G}_{\vec{j}}(\tilde{\rho}_e)}}$  
        \EndFor      
    \EndFor
\EndFor
\end{algorithmic}
where $\M\subset\Natural$ is some choice of sequence lengths, and $(\rho_e,E_E)$
is the SPAM configuration specified by experiment type $e$.
Here, $\RV{\cdot}$ denotes sampling a random variate from 
the given distribution, so that $\RV{J_{M,e}}$ denotes 
choosing a random allowable sequence, and 
$\RV{\Binom{N}{ \Tr \tilde{E}_e \tilde{\G}_{\vec{j}}(\tilde{\rho}_e)}}$
corresponds to repeating this experiment $N$ times and
summing the resulting $0$s and $1$s.
This binomial model assumes strong measurement with two outcomes.
This condition can be loosened, as discussed in 
\autoref{sec:departing-from-bernoulli-trials}.

In principle the number of random sequences $I$ can depend on
$M$ and $e$, and the number 
of repetitions $N$ can depend on $M$, $e$, and $i$, and so on,
but we avoid this to maintain subscriptural sanity (though our
methods will work nonetheless on such ragged structures).
For the same reason, we omit any indices which 
are not relevant to some specific protocol. 
Generically, this protocol produces the dataset
\begin{equation}
    D = (Q_{M,e,i})_{
        M\in\M,
        e\in\Eset_M,
        1\leq i \leq I
    }.
    \label{eq:dataset}
\end{equation}

As a concrete example, consider standard RB.
Then $\Gset$ is a unitary 2-design which is also a group.
There is only one type of experiment for every sequence length,
so $\Eset=\{0\}$, with a fixed SPAM configuration 
$E_0,\rho_0\approx \ketbra{0}$.
We note that our notation allows, however, for formalizing modifications in which two different final measurements are used to decorrelate preparation and measurement errors \cite{fvh+_nonexponential_2015}.
For sequence length $M$ we require $\ell(M,0)=M+1$ gates from 
$\Gset$, where the extra gate corresponds to the final inversion:
the allowable sequences at sequence length $M$ are a uniform distribution
of all length $M+1$ gate indices that ideally produce
the identity gate, 
$J_{M,0}= \Unif{\{\vec{j}\in\{1,...,R\}^{M+1}|\mathcal{G}_{\vec{j}}=\I\}}$.

Interleaved randomized benchmarking has a similar structure except, 
for example, that we may have $\Eset_M=\{0,3\}$, where $e=0$ represents 
no interleaving, and $e=3$ represents interleaving 
the $3^\text{rd}$ gate in $\Gset$.
As with standard RB, we have $\ell(M,0)=M+1$.
Interleaved experiments add a fixed gate for every random 
gate giving us $\ell(M,3)=2M+1$.
See \autoref{tab:models} for more examples of RB+ protocols
as described by our framework.

\section{Tying functions}
\label{sec:survival-distributions}

The quantity
\begin{equation}
    S_{\vec{j},e} = \Tr \tilde{E}_e \tilde{\G}_{\vec{j}}(\tilde{\rho}_e)
        \in [0,1]
\end{equation}
is called the
\textit{survival probability} of the sequence $\vec{j}\sim J_{M,e}$ 
at sequence length $M$ for experiment type $e\in\Eset$.
For a specific noise model and any protocol described by the previous 
section, we can consider the discrete \textit{survival distribution} 
for sequences of length $M$ and experiment type $e$ given by
\begin{equation}
    S_{M,e}(q) = 
        \sum_{\vec{j}} 
            \Pr(\vec{j})\delta(q-S_{\vec{j},e})
    \label{eq:survival-distribution}
\end{equation}
where $\delta(\cdot)$ is the delta mass distribution centered 
at $0$, the sum is over all sequences of the right length,
$\vec{j}\in\{1,...,R\}^{\ell(M,e)}$, and $\Pr(\vec{j})$ is the 
probability of picking sequence $\vec{j}$ according to the protocol.
This distribution has support lying in the unit interval $[0,1]$.

Such survival distributions depend heavily on the noise model.
Complications to the noise model can be introduced successively.
See Epstein et al.~\cite{epstein_investigating_2014} for a wide set of
examples, or Ball et al.~\cite{ball_effect_2016} for simulations of
non-Markovian noise model survival distributions in particular.
Letting $\E_{(\cdot)}$ denote a CPTP noise channel 
and $\vec{j}=(j_1,\ldots,j_K)$ a specific gate sequence, 
starting with the simplest, the broad categories of noise models are
\begin{itemize}
    \item Gate-independent noise: For every $\G_r\in\Gset$ 
        we have $\tilde{\G}_r=\E\G_r$ so that 
        $\tilde{\G}_{\vec{j}}=\E\G_{j_K}\cdots\E\G_{j_1}$.   
    \item Gate-dependent noise: For every $\G_r\in\Gset$ we have
        $\tilde{\G}_r=\E_r\G_r$ so that 
        $\tilde{\G}_{\vec{j}}=\E_{j_K}\G_{j_K}\cdots\E_{j_1}\G_{j_1}$.
    \item Gate- and position- dependent noise:
        For $\G_r\in\Gset$ appearing at time $k$ we have
        $\tilde{\G}_r=\E_{r,k}\G_r$ so that 
        $\tilde{\G}_{\vec{j}}=\E_{K,j_K}\G_{j_K}\cdots\E_{1,j_1}\G_{j_1}$.
\end{itemize}
We can fine-grain these categories further by specifying the types of
channels the errors $\E$ can take, for example, depolarizing, extremal,
or unitary rotations.
We can also, as a matter of preference,
move gate noise to the right side of the ideal operator, 
or consider both left and right noise.
Non-markovian noise models obeying causality are also reasonable to study,
\begin{itemize}
    \item Non-markovian gate dependent noise: 
         For every $\G_r\in\Gset$ we have
        $\tilde{\G}_r=\E\G_r$ where $\E$
        depends on both $r$  and the gates preceding $\G_r$, so that
        $\tilde{\G}_{\vec{j}}=\E_{j_1,...,j_K}\G_{j_K}\cdots\E_{j_2,j_1}\G_{j_2}\E_{j_1}\G_{j_1}$.
\end{itemize}

The set of allowable sequences $J_{M,e}$ typically grows exponentially 
with the sequence length $M$, and numerical evidence suggests that it 
is reasonable to approximate the survival distribution by a continuous 
distribution.

RB+ protocols have the shared property of \textit{tying 
together moments of survival distributions} to extract parameters 
of interest.
For example, the gate-independent noise model ties the first moments
of RB survivals distributions through the relationship\footnote{Recall that 
we omit some indices, here in $S_{M,e}$, for notational convenience. 
In this case because there is only one experiment type and SPAM setting. 
Also, the notation $\mathbb E_{S_M}[q]$ is the expectation value of 
the random variable (arbitrarily called $q$) drawn according to 
the distribution defined by $S_M$ in \autoref{eq:survival-distribution}.}
\begin{equation}
    \expect_{S_M}[q] = (A-B)p^M + B
    \label{eq:rb-zeroth-order}
\end{equation}
where the average gate fidelity of the error map $\E$
is $p+(1-p)/d$, $A=\Tr \tilde{E}_0\E(\tilde{\rho}_0)$, and
$B=\Tr \tilde{E}_0\E(\I/d)$.
Note that we have chosen a slightly different parameterization than that of \citet{magesan_characterizing_2012}, such that the range of valid SPAM parameters is given by $(A, B) \in [0, 1]^2$.

More generally, every protocol will have a function $T$ 
which ties together the $t^\text{th}$ moments of the survival 
distributions through
\begin{equation}
    \expect_{S_{M,e}}[q^t] = T(t, M, e, \vec{x}_T).
\end{equation}
for some subset of all moments.
We call $T$ the \textit{tying function}.
Here, $\vec{x}_T$ is a vector of parameters required by
the tying function, for instance,
$\vec{x}_T=(p,A,B)$ in the case of standard RB.
As of this writing, the unitarity protocol is the only
protocol which ties together moments 
past the first \cite{wallman_estimating_2015}.

\section{The Likelihood Function}
\label{sec:likelihood-function}

\begin{table*}[t]
    \centering
    \scriptsize
    \begin{tabularx}{\textwidth}{lllX}
        \textbf{Protocol} & \textbf{Parameter} & \textbf{Symbol} & \textbf{Value} \\
        \hline \\
        RB~\cite{magesan_characterizing_2012,combes_logical_2017} & Gate Set & $\Gset$ & Group and unitary 2-design, $R$ members \\
         & Experiment Types & $\Eset$ & $\Eset_M=\{0\}$ with SPAM $\rho_0,E_0\approx \ketbra{0}$ \\
         & Allowable Sequences & $J_{M,e}$ & $\Unif{\{\vec{j}\in\{1,...,R\}^{M+1}|\mathcal{G}_{\vec{j}}=\I\}}$ \\
         & Tying Parameters & $\vec{x}_T$ & $(p, A, B)$ \\
         & Tying Functions & $T$ & $T(1,M,e,\vec{x}_T)=(A-B)p^M+B$ \\
        \hline \\
        Interleaved RB~\cite{magesan_efficient_2012} & Gate Set & $\Gset$ & Group and unitary 2-design, $R$ members \\
         & Experiment Types & $\Eset$ & $\Eset_M=\{0,r\}$ for some $1\leq r\leq |\Gset|$, with SPAM $\rho_e,E_e\approx \ketbra{0}$ \\
         & Allowable Sequences & $J_{M,e}$ & $J_{M,0}=\Unif{\{\vec{j}\in\{1,...,R\}^{M+1}|\mathcal{G}_{\vec{j}}=\I\}}$ \\
         &  &  & $J_{M,r}=\Unif{\{\vec{j}\in\{1,...,R\}^{2M+1}|\mathcal{G}_{\vec{j}}=\I,\vec{j}_{\text{even}}=r\}}$ \\
         & Tying Parameters & $\vec{x}_T$ & $(p_0, p_r, A, B)$ \\
         & Tying Functions & $T$ & $T(1,M,e,\vec{x}_T)=(A-B)p_e^M+B$ \\
        \hline \\
        Unitarity~\cite{wallman_estimating_2015} & Gate Set & $\Gset$ & Group and unitary 2-design, $R$ members \\
         & Experiment Types & $\Eset$ & $\Eset_M=\{0\}$ with SPAM $\rho_0,E_0\approx \ketbra{0}$ \\
         & Allowable Sequences & $J_{M,e}$ & $\Unif{\{1,...,R\}^{M}}$ \\
         & Tying Parameters & $\vec{x}_T$ & $(u,A,B)$ \\
         & Tying Functions & $T$ & $T(2,M,\vec{x}_T)=A+Bu^{M-1}$ \\
        \hline \\
        Leakage RB~\cite{wood_quantification_2017} & Gate Set & $\Gset$ & Group and unitary 2-design with $R$ members acting on $\mathcal{\X}_1$, \\
         &  &  & $\X=\X_1\oplus\X_2$ with $\dim\X=d_1+d_2$ \\
         & Experiment Types & $\Eset$ & $\Eset_M=\{0,...,d_1-1\}$ with SPAM $E_e\approx \ketbra{e},\rho_e \approx \ketbra{0}$ \\
         & Allowable Sequences & $J_{M,e}$ & $\Unif{\{\vec{j}\in\{1,...,R\}^{M+1}|\mathcal{G}_{\vec{j}}=\I\}}$ \\
         & Tying Parameters & $\vec{x}_T$ & $(
    L_1,
    L_2,
    \mu_1,
    p_l,
    \{A_e\}_{e\in\Eset},
    \{B_e\}_{e\in\Eset},
    \{C_e\}_{e\in\Eset},
)$ \\
         & Tying Functions & $T$ & $T(1,M,e,\vec{x}_T) =$$(L_2 A_e + L_1 B_e)/(L_1+L_2)$ \\
         &  &  & $\quad\quad+\left(\frac{L_1}{L_1+L_2}-p_l\right)$$(A_e-B_e)(1-L_1-L_2)^M$ \\
         &  &  & $\quad\quad+(1-p_l)(C_{e}-A_e)(\mu_1(1-L_1))^M$ \\
        \hline \\
        Dihedral Benchmarking~\cite{carignan-dugas_characterizing_2015} & Gate Set & $\Gset$ & $\langle Z_j=\ee^{\ii \pi Z/j},X\rangle\subseteq \SU(2)$ for some $j\in\Natural$, $R$ total members \\
         & Experiment Types & $\Eset$ &
         $\Eset_M=\{X,Z\}$ with SPAM $E_e, \rho_e\approx (I+e)/2$ \\
         & Allowable Sequences & $J_{M,e}$ &
         $\Unif{\left\{\vec{j}\in\{1,...,R\}^{M+1}|\mathcal{G}_{\vec{j}} \in
         \{I,e\}
    \right\}}$ \\
         & Tying Parameters & $\vec{x}_T$ & $(p_X,p_Z,A,B_X,B_Z)$ \\
         & Tying Functions & $T$ & $T(1,M,e,\vec{x}_T)= A + B_e p_e^M$ \\
        \hline \\
    \end{tabularx}
    \caption{
Description of some RB+ protocols within our framework.
}
    \label{tab:models}
\end{table*}

Let's start with the standard RB protocol in what is known as 
the $0^\text{th}$ order model, as written in \autoref{eq:rb-zeroth-order}.
The parameter of interest is $p$ since it is related to the average gate 
fidelity of the average error map.
Given a dataset $D$, as defined in \autoref{eq:dataset}, we are 
interested in inferring the value of $p$, with $A$ and $B$ treated 
as nuisances.

Any inference starts with writing down the likelihood function of the
parameter of interest \cite{ste_bayesian_2007}, along with nuisance parameters, conditioned 
on the collected data.
The total likelihood will be a product over all sequences 
lengths and sequence draws.
Consider just the factor for  the $i^\text{th}$ draw 
of length-$M$,
resulting in the binomial outcome $d=Q_{M,e,i}\in\{0,\ldots,N\}$.
The likelihood of this outcome, conditional on 
drawing the particular sequence $\vec{j}$, is given by
\begin{equation}
    \lhood(p,A,B|d,\vec{j})
        = \binom{N}{d} q^d(1-q)^{N-d}
\end{equation}
where $q=S_{\vec{j}}$ is the survival probability of sequence $\vec{j}$.
The conditional is removed by marginalizing $q$ over the survival
distribution,
\begin{equation}
    \lhood(p,A,B|d,M)
        = \expect_{S_{M}}\left[
              \binom{N}{d} q^d(1-q)^{N-d}
           \right].
    \label{eq:likelihood-exact}
\end{equation}
At this point we have run into a very serious problem.
This expression cannot be simplified, even in principle, unless 
we know more about the survival distribution $S_{M}$.
There is one exception, however, first explicitly pointed 
out in an appendix of Granade et al.~\cite{granade_accelerated_2015}: 
if $N=1$, then the expectation's 
integrand is linear in $q$ for both values of $d$ 
and so only the first moment of $S_{M}$ matters; we get
\begin{align}
    \lhood(p,A,B|d=1,M)
        = (A-B)p^M+B
\end{align}
for standard RB, or more generally,
\begin{align}
    \lhood(\vec{x}_T|d=1,M,e)
        = T(1,M,e,\vec{x}_T)
\end{align}
for any protocol whose first moments are tied together.
This fact was exploited to great effect by those authors.
The same argument shows that the first $N$ 
moments of $S_{M,e}$ are potentially relevant to the likelihood
function for any protocol, and therefore some characterization of them should 
be appended to the list of nuisance parameters.

Alternatively, one might argue to simply enforce the constraint 
$N=1$.
This is a reasonable suggestion, and is explored in
\autoref{sec:sequence-re-use} where it is shown that
$N=1$ should be considered best-practice for 
protocols which only tie together their first moments, and whose
implementations are quick at switching between random sequences.
For some experimental setups, however, switching the sequence
every experiment would dominate the duty cycle.
The way around this is through fast logic near the quantum
system \cite{rjr+_hardware_2017}, such as was recently demonstrated by \citet{hro+_implementing_2016} in the case of a transmon qubit coupled to an oscillator-encoded logical qubit.
Or perhaps, even more seriously, some systems are not capable of
strong measurement, and so a binomial model with $N=1$ is not
physically possible.
In this case we can still write down a likelihood function,
no longer conditionally binomial as seen in 
\autoref{sec:departing-from-bernoulli-trials}, but one that 
will involve higher moments by necessity.
Finally, in some cases, the second moment is the moment of interest,
as in the unitarity protocol, so that $N=1$ is completely 
insensitive to the quantity of interest. 

In any case, a great deal of RB+ experiments have been performed with $N>1$ and
so it behooves us to devise a statistically rigorous approach for analysing such
data.

\section{Constructing Agnostic Models}
\label{sec:constructing-agnostic-models}

In the last section we noted that for a repetition value of $N$, to
fully specify the likelihood function of an RB or related protocol,
we require at least $N$ parameters per sequence length and experiment type, 
in addition to the parameters of the tying function.
These extra parameters correspond to moments of the survival distributions.
We will write $\vec{x}_S$ to denote these new parameters, whatever they end 
up being, distinguishing them from the parameters of the tying function,
$\vec{x}_T$.
One must tread carefully in any analysis that follows this observation.
The goal of this section to develop a framework where we treat these 
nuisance parameters in a principled yet practical way, while at the same
time remaining as agnostic about their structure as possible.

\subsection{Parameterizations}

A Bayesian, by instinct, may be tempted to throw all of the unknown
moments of the survival distributions into an inference 
engine as nuisance hyperparameters.
In principle there is nothing wrong with this.
However, it would lead to a huge number 
of parameters for even modest values of $N$.
Care would be required in restricting the domains of these moments,
for example, the variance $\sigma^2$ of a distribution with 
support on $[0,1]$ and expectation value $\mu$ must always satisfy 
$0\leq \sigma^2 \leq \mu(1-\mu)$.

One might suggest next to truncate the number of moments to be
included as hyperparameters down 
to some tractable, empirically motivated constant. 
But even in this case, one must specify the higher moments somehow. 
For example, one might choose to set them all to zero.
This would effectively restrict the space of 
allowed survival distributions to some strange, unmotivated family of distributions.
Instead, one might make the moments above the truncation cutoff sure 
functions of those below in some sensible way.

At this point, we have basically argued for the use of parameterized
families of probability distributions; any family of probability
distributions, like the Gaussian or gamma families, can be defined as a rule 
that specifies all moments of a given member in terms of a few parameters.
For us, the most natural starting point is the beta distribution family.
This family is conjugate to the binomial distribution, and
is the canonical family of continuous distributions with support on the unit interval.
A member with parameters $\alpha,\beta>0$ is written $\betadist(\alpha,\beta)$, 
and has a density function defined by
\begin{equation}
    \pdf{\betadist}{q} = \frac{q^{\alpha-1}(1-q)^{\beta-1}}{\betafun{\alpha,\beta}}
    \label{eq:beta-pdf}
\end{equation}
where the normalization constant $\betafun{\alpha,\beta}$ is the beta function.
Its first and second central moments are given by $\mu=\frac{\alpha}{\alpha+\beta}$
and $\sigma^2=\frac{\alpha\beta}{(\alpha+\beta)^2(\alpha+\beta+1)}$, respectively.
These equations can be uniquely inverted as
\begin{subequations}
    \begin{align}
	    \alpha &= \mu^2(1-\mu)/\sigma^2-\mu \\
	    \beta  &= \mu(1-\mu)^2/\sigma^2-(1-\mu),
    \end{align}
    \label{eq:beta-param}
\end{subequations}
which provides an alternate parameterization of the family.
In a slight abuse of notation, we write $\betadist(\mu,\sigma)$ for a member 
written in the new coordinates.
Alternate parameterizations and their transforms are provided in 
\autoref{app:dist-reparam}, and we similarly 
abuse notation for these other coordinates, writing, for example,
$\betadist(\mu,r)$ where $\sigma^2=r\mu^2(1-\mu)^2$.

This family can produce quite a wide variety of shapes even though it only
has two parameters.
Setting $\alpha=\beta=1$ results in the uniform distribution on $[0,1]$.
Fixing any mean $\frac{\alpha}{\alpha+\beta}\in(0,1)$ while increasing 
$\alpha$ and $\beta$ decreases the variance, and the distribution approaches a
normal shape.
On the other hand, decreasing $\alpha$ and $\beta$ while the mean
is kept fixed increases the variance toward $\mu(1-\mu)$; the probability density
at first spreads out over the whole interval $[0,1]$, and when this is no 
longer able to keep increasing the variance, the mass begins to 
build up at the end points, approaching a weighted mixture of two delta functions.

Using this family, for a first order tying function,
every sequence length, experiment 
type, and measurement operator would add one parameter
 to the likelihood model, so that
$\vec{x}_S=\{\sigma_{M,e}\}_{M\in\M,e\in\Eset_M}$,
or some other parameterization thereof.
In the case of any protocol which ties together only first moments,
we get the hierarchical model
\begin{subequations}
    \begin{align}
        \vec{x}_T &\sim \pi(\vec{x}_T) \\
        \mu_{M,e}|\vec{x}_T &= T(1,M,e,\vec{x}_T) \\
        \sigma_{M,e} &\sim \pi(\sigma_{M,e}) \\
        \mathclap{\rule{5cm}{0.2pt}} \nonumber \\
        q_{M,e,i} | \mu_{M,e},\sigma_{M,e} 
            &\iid \betadist\left(\mu_{M,e}, \sigma_{M,e}\right) \\
        Q_{M,e,i} | q_{M,e,i} &\iid \Binom{N}{q_{M,e,i}}
    \end{align}
    \label{eq:betabin-model}
\end{subequations}
for the dataset $D$.
The horizontal line is a visual aid to separate the prior 
from the likelihood distribution, and $\pi(\cdot)$ 
refers to the prior distribution of the given parameters.
The quantities $q_{M,e,i}$ are latent random variables representing
survival probabilities ---
they can be analytically integrated out of the model if desired, resulting
in a beta-binomial distribution instead.
This set of sampling statements, which are sequentially 
dependent on previous variables, is an example of a probabilistic program.
It is a convenient way of specifying the joint distribution of the
prior and the likelihood, which is proportional to the 
posterior distribution.

Models for higher-order tying functions are just as easy to write down.
Note, however, that the beta distribution only has two parameters, so
that if both of the first two moments are tied together, there is no more 
uncertainty in the survival distributions (conditional on a specific value
of $\vec{x}_T$).
This can be solved by using a larger family of distributions, or
through a nonparametric approach, as discussed in the following section.

\subsection{Nonparameterizations}

The assertion that every survival distribution is approximately 
beta distributed may sometimes be too strong.
In this section we would like to loosen this restriction.
One viable path is to use a bigger family, such as the generalized
beta family with five parameters~\cite{mcdonald_generalization_1995}.
Even more generally, we can resort to Bayesian nonparametrics.
This is the approach that we take, and in particular, we use 
Dirichlet process mixtures, which 
are distributions of distributions\footnote{We provide a brief 
introduction to Dirichlet processes and Dirichlet process mixtures in 
\autoref{app:dirichlet-process}.}.

Let $\dirp_K(\alpha, G_0)$ denote a Dirichlet process with a concentration 
parameter $\alpha>0$ and a base distribution $G_0$ that has support on the 
parameter space $\Omega$, and that is truncated to $K$ modes\footnote{As a
brief bit of context, recall that $G_0$ is the mean value 
of $\dirp_K(\alpha, G_0)$, and that $\alpha$ can be interpreted 
as the number of `prior observations' from samples of $\dirp_K(\alpha, G_0)$; 
it scales inversely with the variance of $\dirp_K(\alpha, G_0)$.}.
We would like to replace the draw of $q_{M,e,i}$ from a beta distribution 
(see \autoref{eq:betabin-model})
to a draw from a random distribution $G$, such as $G\sim\dirp_K(\alpha, G_0)$.
The Dirichlet process has two shortcomings that prevent us from directly 
using it for this purpose.
The first is that its variates are not continuous distributions, and 
the second is that the moments of its draws are random, whereas
we would like the ability to (conditionally) fix some of them according to the
tying functions.

To overcome these problems we modify the Dirichlet process 
into a new nonparametric family that we call
\textit{constrained Dirichlet process beta mixtures} ($\CDPBM$),
denoted $\CDPBM_K(\alpha, G_0, \cdot)$,
whose definition is motivated in \autoref{app:cdpbm-motivation}.
In short, if the desired mean value of our random distributions 
is $0<\mu_1<1$, then the random distribution 
$G\sim\CDPBM_K(\alpha, G_0, \mu_1)$ is drawn as follows:
\begin{subequations}
    \begin{align}
        \sum_{k=1}^{K}w_k \delta_{(\nu_k^*,r_k)} 
            &\sim \dirp_K\left(\alpha, G_0\right) \\
        \nu_k
            &= \frac{1}{1+\ee^{-\nu_k^*-h}} 
                \text{ with }h\text{ such that } 
                \sum_{k=1}^K w_k \nu_k=\mu_1 \\
        G 
            &= \sum_{k=1}^K w_k \betadist(\nu_k, r_k).       
    \end{align}
    \label{eq:cdpbm}
\end{subequations}
Here, the Dirichlet process sample space is 
$(\nu_k^*,r_k)\in\Real \times (0,1)=\Omega$, upon which the base
distribution $G_0$ is defined, and we are using the $(\mu,r)$ 
parameterization of the beta family (see \autoref{app:dist-reparam}).
This procedure ensures that $\expect[G]=\mu_1$, and that the 
support of $G$ lies within $[0,1]$.
We typically choose $G_0=\normal(0,1.9)\times\uniform(0,1)$ as a broad prior,
and assign a hyper-prior $\alpha~\gammadist(1,1)$.

With this defined, our nonparametric model for analyzing RB+ data 
is a straight-forward modification of \autoref{eq:betabin-model},
given by
\begin{widetext}
    \begin{subequations}
        \begin{align}
            \vec{x}_T 
                &\sim \pi(\vec{x}_T) \\
            \mu_{M,e}|\vec{x}_T 
                &= T(1,M,e,\vec{x}_T) \\
            \alpha_{M,e} 
                &\iid \gammadist(1,1) \\
            G_{M,e}|\alpha_{M,e},\mu_{M,e} 
                &\ind \CDPBM_K\left(
                        \alpha_{M,e}, 
                        G_0,
                        \mu_{M,e}
                    \right) \\
            \mathclap{\rule{7.5cm}{0.2pt}} \nonumber \\
            q_{M,e,i}|G_{M,e} 
                &\ind G_{M,e} \\
            Q_{M,e,i}|q_{M,e,i} 
                &\ind \binomial(N, q_{m,i}).
        \end{align}
        \label{eq:dp-first-moment}
    \end{subequations}
\end{widetext}
A slight modification is needed for protocols which tie together higher
moments, which we omit for brevity; see \autoref{app:cdpbm-motivation}.

\subsection{Frequentist Approaches}
\label{sec:frequentist-approach}

Though we are primarily concerned with a Bayesian approach, we are also
interested in comparing to frequentist methods.
To date, the \textit{de facto} frequentist inference tool for RB+ data 
(with exceptions)
has been least-squares fitting (LSF) to exponential decay models.
Generally, the justification for LSF is that it is
equal to the maximum likelihood estimator (MLE) 
in the case of Gaussian noise on the data.

There are a couple of reasons to be cautious when using 
estimates and confidence regions based on LSF in the case of RB+.
One is that the distribution of the data is not Gaussian, except
approximately 
in the high data regime, and therefore the MLE is not being reported,
but some sort of approximation thereof.
Another is that weights need to be chosen for weighted LSF 
(WSLF)---using uniform weights implicitly makes assumptions about
the nature of the noise model and should always be avoided.

It is non-trivial to choose appropriate weights for WLSF.
One may be tempted to use sample variances as weights, 
but there is a subtle issue that these variances do 
not directly represent the uncertainty of the quantities of interest
at a given sequence length and experiment type;
they partially contain unnecessary weight due to finite sampling statistics.
Even if this is corrected for, one must also make sure that weights
are assigned consistently. 
Additionally, one needs a heuristic for assigning a non-zero weight in 
the case of no variance in the outcomes at a given sequence length of
a protocol.

For these reasons our preferred frequentist method for analyzing
data from RB+ models containing 
many sequence lengths is to 
look directly at the MLE.
This can be done by using a likelihood function that assumes
that survival distributions are beta distributed---see the
second half of \autoref{eq:betabin-model}.
The log-likelihood of this model is easily and reliably 
maximized with gradient-based numerical methods.
We avoid having to assign weights at every sequence length since
they are now treated as nuisances of the global fit.
Confidence intervals for this estimator can be constructed through
standard bootstrapping techniques (see for example the survey
article of DiCiccio and Efron~\cite{diciccio_bootstrap_1996}).
In this paper we construct bootstrap distributions 
of the tying parameters by computing
the MLE on random data replications drawn from the 
empirical (non-parametric) distribution of the data, or by
sampling the likelihood distribution at the MLE of the data (parametric).
Samples are always drawn on a per-sequence-length basis, so that 
the shape of the bootstrapped data is the same as that of the original data.
Confidence intervals are constructed with the simplest 
bootstrap-$t$ procedure. That is, we look directly
at the CDF of these bootstrap distributions.

Occasionally we will also consider the WLSF for the sake of interest.
In such cases, we set weights equal to the sample variances 
of the binomial data normalized by $N$. We do this because it 
has been a popular approach historically.

\section{Sequence Re-Use}
\label{sec:sequence-re-use}

Thus far we have only talked about data analysis.
In this section we discuss which experiments to perform 
in the first place.
Specifically, we address the question of how many times a fixed 
random sequence from an RB+ protocol should be reused.
In \autoref{sec:likelihood-function} we hinted at the fact that
every random sequence should, ideally, only be used once.
Here, we qualify and quantify this idea. 

With all of the heavy lifting of getting to the survival distribution out of the way, we can cast the problem of sequence re-use as one of pure statistics. Or, we can think of a concrete and conceptually simple isomorphic problem---we can think of a survival distribution as a bag of coins with different 
biases.
Suppose this bag has a mean bias $\overline{q}$ and a standard deviation 
of biases $\sigma$ (or, equivalently characterized by the second moment $\mu_2$). We want to estimate these unknown quantities from selecting coins from the bag, at random, and flipping them. The isomorphism is that the statistical conclusions of flipping the same coin more than once are the same as repeating a given gate sequence in RB.

So, by considering the trade-off in the number of repetitions of flips using the same coin versus selecting a new coin, we can understand the optimal experimental design policy in RB.

\subsection{First moment estimators}

\begin{figure*}[t]
    \includegraphics[width=\textwidth]{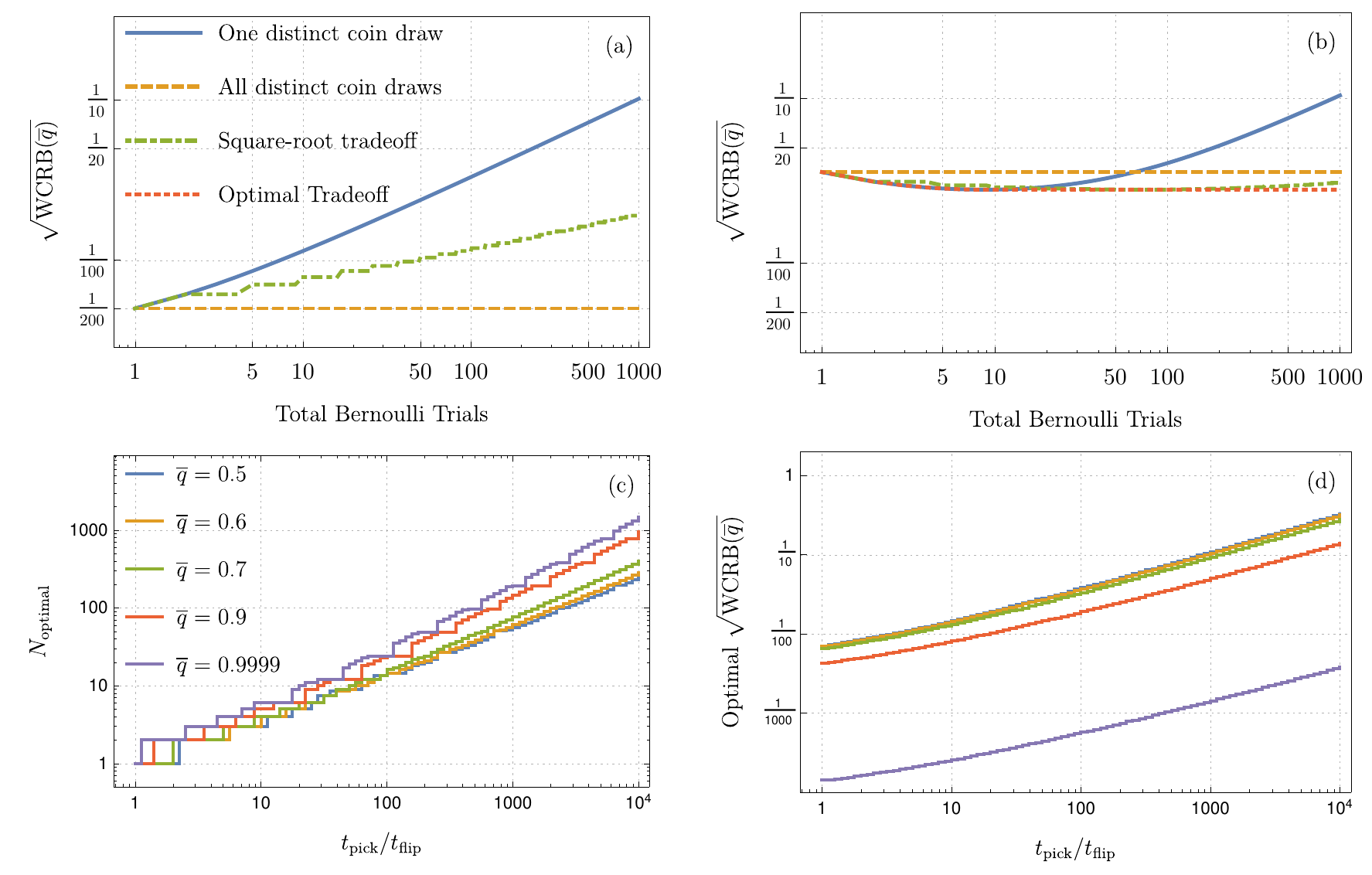}
    \caption{
        (a-b) Supposing a fixed budget of Bernoulli trials for the bag-of-coins
        experiment, the WCRB (\autoref{eq:weighted-crb}) 
        of the mean coin bias $\overline{q}$ is shown, 
        normalized to the time it takes to complete the full experiment.
        The true parameters of the bag are $\overline{q}=t=0.5$, a single
       coin flip takes $t_\text{flip}=\SI{100}{us}$, and switching coins 
       takes $t_\text{pick}=0$ and $t_\text{pick}=\SI{5}{ms} = 50 t_{\text{flip}}$
       for (a) and (b), respectively.
       We see choosing all coins to be different
       is no longer the best strategy when $t_\text{pick}>0$.
       To explore this, in (c-d),
       given a ratio $t_\text{pick}/t_\text{flip}$,
       we compute the optimal number of coin flips $N$ and the resulting
       optimal WCRB for $t_\text{flip}=\SI{100}{us}$, $t=0.5$, and 
       various values of $\overline{q}$.
       (In these final two plots, there is no longer a fixed budget of trials;
       global minima were found with respect to $N$.)
    }
    \label{fig:betabin-crb}
\end{figure*}

Naturally, we start with the first moment. With protocols that tie only first moments, we only care, by necessity, about inferring values which 
depend on $\overline{q}$, but none of the higher moments of the bag.
Conditional on picking a coin with bias $q$, if we perform $N$ Bernoulli
trials and add them up, we have the conditional random variable
\begin{align}
    Q|q \sim \binomial(N, p)
\end{align}
with conditional cumulants $\expect[Q|q]=Np$ and $\variance[Q|q]=Nq(1-q)$.

This gives
$\variance[Q]=N\overline{q}(1-\overline{q})+N(N-1)\sigma^2$
through the law of total variance.
If we independently and identically 
repeat the process of drawing a different coin $I$ times 
and perform $N$ Bernoulli trials on each, we end up with
\begin{align}
    \variance\left[\sum_{i=1}^I \frac{Q_i}{N\cdot I}\right]
        = \frac{1}{I}\left( 
            \frac{\overline{q}(1-\overline{q})
                  }{N} 
            + \frac{N-1}{N}\sigma^2 
        \right)
\end{align}
as the variance of the scaled quantity $\sum_{i=1}^I Q_i/(N\cdot I)$
whose mean value is $\bar{q}$.
The take-away from this formula is that the variance approaches 
$0$ as we increase the number of coins (sequences) we use,
but asymptotes to the finite value $\sigma^2$
if we fix $I$ and increase the binomial parameter $N$ (re-use
of the same sequence).
If we consider instead the total number of flips of all coins to be fixed, $I\cdot N={\rm const.}$, we can see at once that the variance is minimized when $N=1$ by completely eliminating the contribution from $\sigma$.

We have looked at the variance formula above because it has a simple derivation
and gets the point across.
However, a better quantity to consider is the Fisher information and the 
resulting Cram\'er--Rao bound of $\overline{q}$, because it gives a rigorous
bound on how well any (unbiased) estimator of $\overline{q}$ can do.
Supposing that we explicitly choose our bag to have a beta distribution with
mean value $\overline{q}$ and variance $\sigma^2=t\overline{q}(1-\overline{q})$
for some $0<t<1$, then our likelihood distribution is $\betabin(N,\mu,t)$ 
and the two-by-two Fisher information matrix, $J(\overline{q},t)$, 
is given by the negative expected value of the Hessian of the log-likelihood function.
By virtue of our choice of parameterization $(\overline{q},t)$, the 
Fisher information matrix happens to be diagonal, and so the
the Cramer-Rao bound reads
\begin{align}
    \Var[\hat{\overline{q}}] \geq \frac{1}{I\cdot J(\overline{q})}
\end{align}
where $J(\overline{q})=J(\overline{q},t)_{1,1}$ and
$\hat{\overline{q}}(Q_1,...,Q_I)$ is any unbiased estimator of $\overline{q}$ 
that depends on $I$ iid samples from the likelihood.

So far we have neglected any cost associated with picking a new 
coin from our analysis, which is the main reason why experimentalists 
re-use sequences.
We can include this cost by considering the Fisher information per unit 
time, $J(\overline{q})/T$, where $T$ is the time it takes to 
collect the data.
Suppose that it takes time $t_\text{pick}$ to pick a new coin 
and time $t_\text{flip}$ to flip a coin once.
Then we have
$T=I(t_\text{pick}+N t_\text{flip})$, and the CRB 
weighted by experiment cost is
\begin{align}
    \Var[\hat{\overline{q}}]/\text{Hz}
       &\geq \frac{(t_\text{pick}+N t_\text{flip})}{J(\overline{q})}
       \equiv \operatorname{WCRB}(\overline{q})
   \label{eq:weighted-crb}
\end{align}
where we have assumed $T$ is in units of seconds.
Note that if we take the square root of both sides we get the usual
units for sensitivity.
This figure of merit is explored in \autoref{fig:betabin-crb}.

\subsection{Second moment estimators}

As before, we draw a coin $I$ times and perform $N$ Bernoulli trials on each. This time, however, we estimate the \emph{second moment} via summing the squares of the number of successes. This estimator is biased, but not asymptotically so:
\begin{equation}
    \expect\left[\sum_{i=1}^I \frac{Q_i^2}{I\cdot N^2}\right] = \mu_2 + \frac1N(\overline{q} - \mu_2).
\end{equation}
That is, as the number of repetitions $N$ increases, this estimator becomes less biased.

Due to this bias, the Cram\'er--Rao cannot tell us much about this estimator. But, we can directly calculate the mean squared error. As before, though, we consider a fixed total number of measurements $T = N\cdot I =$ const. and calculate $T\cdot$ MSE.

Since the MSE involves the square of the second moment, we need to calculate
\begin{align}
    &\expect\left[\left(\sum_{i=1}^I \frac{Q_i^2}{I\cdot N^2}\right)^2\right]  = \frac{1}{I^2N^4}\sum_{j,k=1}^I \expect[Q_j^2 Q_k^2] \nonumber \\
    & = \frac{1}{I^2N^4} \left(\sum_{k=1}^I \expect[Q_k^4] +\sum_{j\neq k = 1}^I \expect[Q_j^2]\expect[Q_k^2]\right) \nonumber \\
    & = \frac{1}{I^2N^4} \left(I \expect[Q^4] +I(I-1) \expect[Q^2]^2\right).
\end{align}
The fourth moment of the Beta-Binomial $\betabin(N,\mu,\mu_2)$ is simple yet still too messy to usefully reproduce here.

We calculate the optimal repetition rate by averaging the total cost over a uniform prior on the domain of validity in the parameterization of $(\mu,\mu_2)$.
The final answer for the optimal value of $N$ is
\begin{equation}
    \label{eq:unitarity-opt-reuse}
    N_{\rm opt} = \left(\frac{16}{40+32\ln(2)-3\ln(3)}\right)^{\frac13} T^{\frac13}+ O\left(\frac{1}{T^{\frac13}}\right),
\end{equation}
or roughly $0.65 T^\frac13$. A ball-park amount of data usually taken at each sequence length in randomized benchmarking is about a kilobyte.
This corresponds to about $N = 13$ repetitions per sequence and $I = 615$ difference sequences.

It is also of interest to consider the case when $\mu \in (l,1)$ for some lower bound $l$. 
For example, suppose we are fairly confident that our fidelity is above $90$\%. 
In this case, we still have
\begin{equation}
    N = C(l) T^{\frac13}+ O\left(\frac{1}{T^{\frac13}}\right),
\end{equation}
for some $C(l) < C(0)$. 
For example, taking $l = 0.9$, we have $N = 0.39 T^{\frac13}$.

\begin{figure}
    \begin{centering}
        \includegraphics[width=0.5\columnwidth]{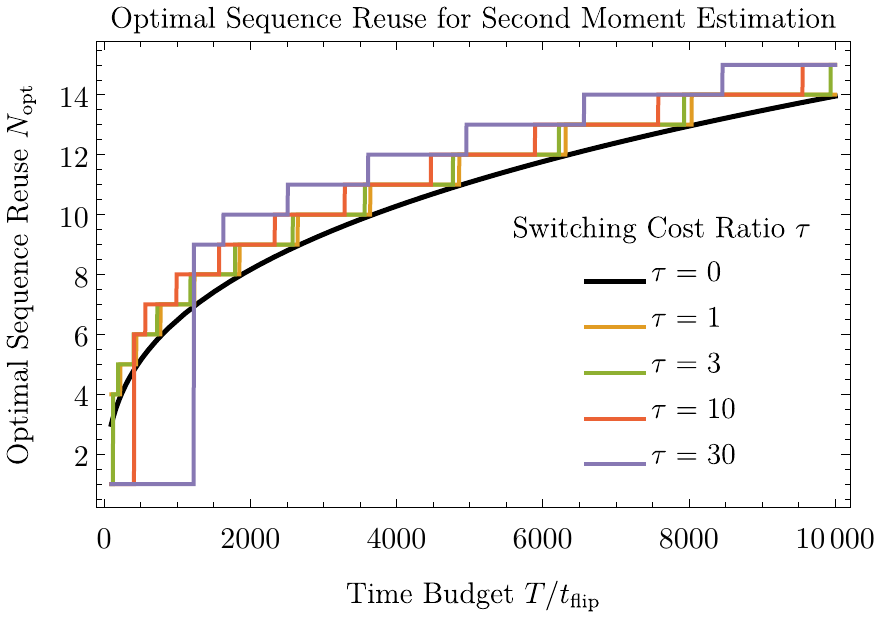}
    \end{centering}
    \caption{
        \label{fig:unitarity-optimal-switching-cost}
        The optimal sequence reuse $N_{\text{opt}}$ for second
        moment estimation (as used, for example, in the unitarity 
        protocol), plotted as a function of the total time budget 
        allowed $T$, for each of several choices of the 
        switching cost ratio 
        $\tau \defeq t_{\text{pick}} / t_{\text{flip}}$.
    }
\end{figure}

Finally, we generalize the calculation of \autoref{eq:unitarity-opt-reuse} to include the effects of finite switching costs $\tau \defeq t_{\text{pick}} / t_{\text{flip}}$.
In doing so, we proceed numerically, as the series expansion obtained in \autoref{eq:unitarity-opt-reuse} is much less useful for $\tau > 0$.
We plot the results in \autoref{fig:unitarity-optimal-switching-cost}, noting that even for $\tau = 30$, the optimal sequence lengths found do not deviate substantially from the case where there is no switching cost.
Thus, $N_{\mathrm{opt}} \approx 0.65 T^{\frac13}$ remains a useful heuristic in this case, even if it is no longer a rigorous approximation.

\section{Numerical Results}
\label{sec:numerical-results}

In this section we explore our Bayesian model with a collection of 
numerical examples, using various protocols and error models.
Code to reproduce these results can be found online~\cite{code_repo}.

As with most Bayesian models, 
analytic formulae for posterior distributions are intractable.
Our posterior in the examples throughout this section are therefore
computed with numerical techniques.
In particular, we use the Hybrid Monte Carlo (HMC) sampler using the No-U-Turns (NUTS) heuristic \cite{duane_hybrid_1987,hoffman_no-u-turn_2014}.
This is a type of Markov chain Monte Carlo (MCMC) sampler 
that has gained widespread use due to its lack of tuning parameters, 
fast mixing rate, and ability to handle large numbers of parameters.
More details about our sampling strategies are outlined in 
\autoref{app:sampling-strategies}.

\subsection{RB with Various Noise Models}
\label{sec:rb-various-noise-models}

\begin{figure*}[t]
    \includegraphics[width=\textwidth]{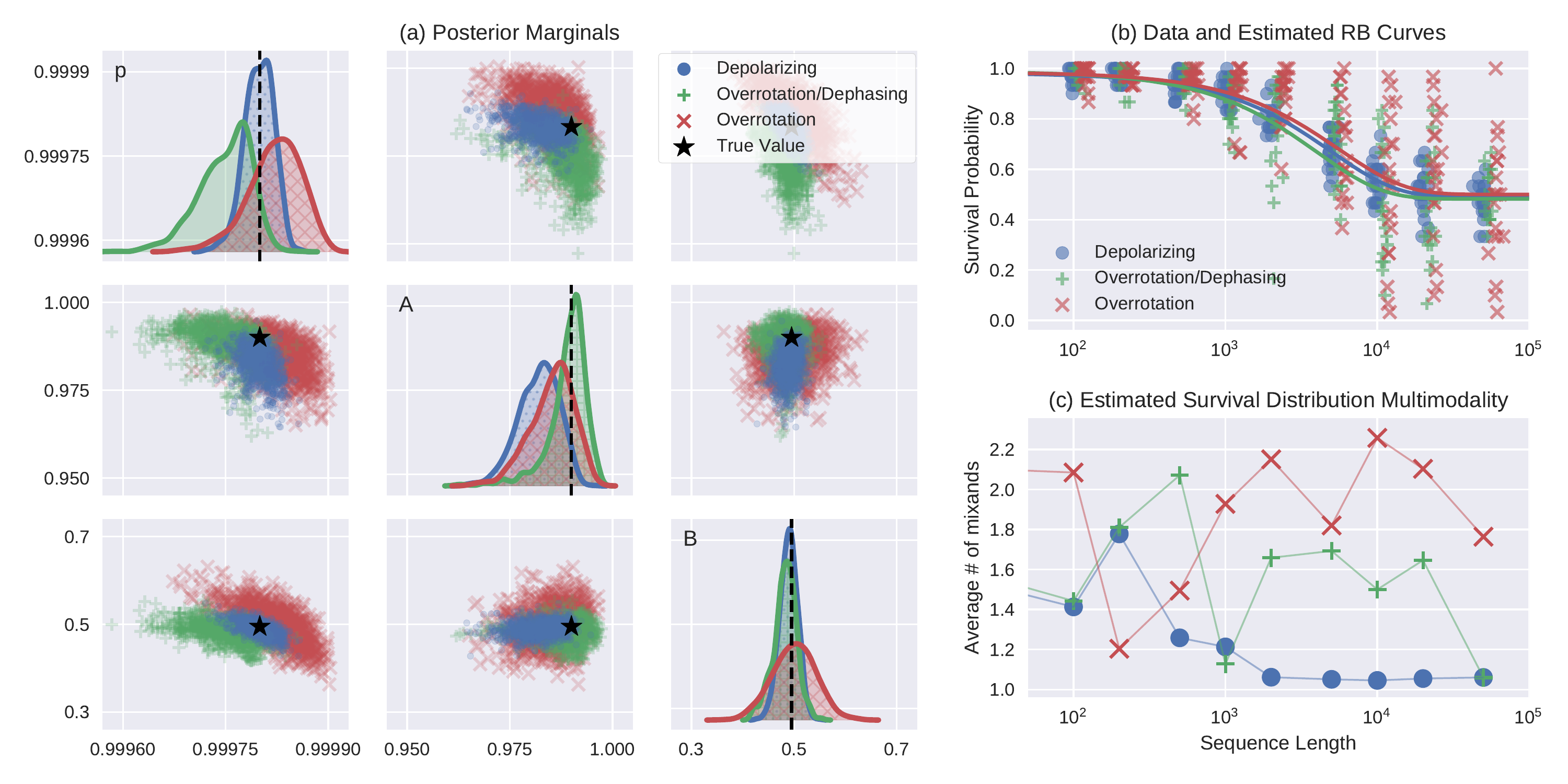}
    \caption{
    (a) Single and joint posterior marginals 
    of the parameters $p$, $A$, and $B$ are shown for each of the 
    three noise models defined in \autoref{sec:rb-various-noise-models}
    of the main text for the standard RB protocol.
    (b) Using Bayes' estimate for these three parameters, the curve
    $(A-B)p^M+B$ is plotted for each model, displayed on top of the
    normalized data used in the inference.
    The unusual shape is due to the log-linear scale, and jitter in
    the $x$-axis on the data points was added for visual appeal ---
    for all three models $I=20$ random sequences were used with 
    $N=30$ repetitions each at each of the sequence lengths 
    $\M=\{1,100,200,500,1000,2000,5000,10000,20000,50000\}$,
    where the maximum sequence length is chosen according to the $M_{\max} = 1 / (1 - F)$ heuristic \cite{granade_accelerated_2015}.
    (c) The posterior shown in (a) was calculated using the model in 
    \autoref{eq:dp-first-moment}, which describes each survival 
    distribution as a mixture of beta distributions, and so finally,
    we plot the posterior mean of $1/\sum_{k=1}^K{w_k^2}$ for 
    each survival distribution, where the 
    weights $w_k$ are defined in \autoref{eq:cdpbm}.
    This quantity ranges between $1$ and $K$ and quantifies the 
    estimated number of relevant mixands in each survival distribution.
    The low values justify our CDPBM truncation at $K=10$.
    }
    \label{fig:different-noise-models-summary}
\end{figure*}

As a first example, we consider the standard RB protocol on 
a qubit under three noise models.
We use an order 12 subgroup of the usual 24 member
Clifford group as our gateset. 
This subgroup is still a 2-design and can be 
generated as $\Gset=\langle Z, \sqrt{Z}H\rangle$,
where $H=\inlinetwobytwo{1}{1}{1}{-1}/\sqrt{2}$ and 
$Z=\inlinetwobytwo{1}{0}{0}{-1}$.
Our three noise models are defined as
\begin{subequations}
\begin{align}
    \E_r^1 &= \Lambda_{s_1} \\
    \E_r^2 &= \Phi_{s_2}\circ\Theta[G_r,\epsilon_2] \\
    \E_r^3 &= \Theta[G_r,\epsilon_3]
    \label{eq:overrotation-model}
\end{align}
\end{subequations}
where $\tilde{\G}_r=\G_r\circ\E_r^i$ is the actual implementation 
of the ideal gate $\G_r$ for $r=1,...,R$ and where
\begin{subequations}
\begin{align}
    \Lambda_{s}(\rho) &= (1-s)\rho+s\Tr[\rho]\I/2 \\
    \Phi_{s}(\rho) &= (1-s)\rho+sZ \rho Z \\
    \Theta[U,\epsilon](\rho) &= \begin{cases}
            \rho & U\text{ is some z-rotation} \\
            U^{\epsilon}\rho(U^{\epsilon})^\dagger & \text{else}
        \end{cases}
\end{align}
\end{subequations}
are the depolarizing, dephasing, and transverse overrotation channels,
respectively.
Therefore $\E_r^1$ is a gate independent depolarizing channel,
$\E_r^2$ is gate independent dephasing combined with a gate dependent
overrotation by amount $\epsilon_2$,
and $\E_r^3$ is purely gate dependent overrotation by amount $\epsilon_3$.
Constants were chosen by trial and error 
so that all three noise models result in exactly
the same RB decay base $p=0.9998$, ultimately achieved with 
the choices $s_1=0.0002$, $s_2=0.000028954$, $\epsilon_2=0.01$, and
$\epsilon_3=0.11132$.
A formula for computing $p$ given a gate dependent noise model is 
provided in Ref.~\cite{wallman_randomized_2017}.

Data was simulated under each noise model 
with the initial state $\rho=\ketbra{0}$ and 
the measurement $M=0.99\ketbra{0}$ at each of the sequence lengths
$\M=\{1,100,200,500,1000,2000,5000,\allowbreak 10000,20000,50000\}$.
At each sequence length, $I=20$ random sequences were drawn and 
$N=30$ repetitions were used for each.
To produce histograms of the survival distributions, however,
thousands of simulations were done per sequence length.

\begin{figure*}[!ht]
    \includegraphics[width=\textwidth]{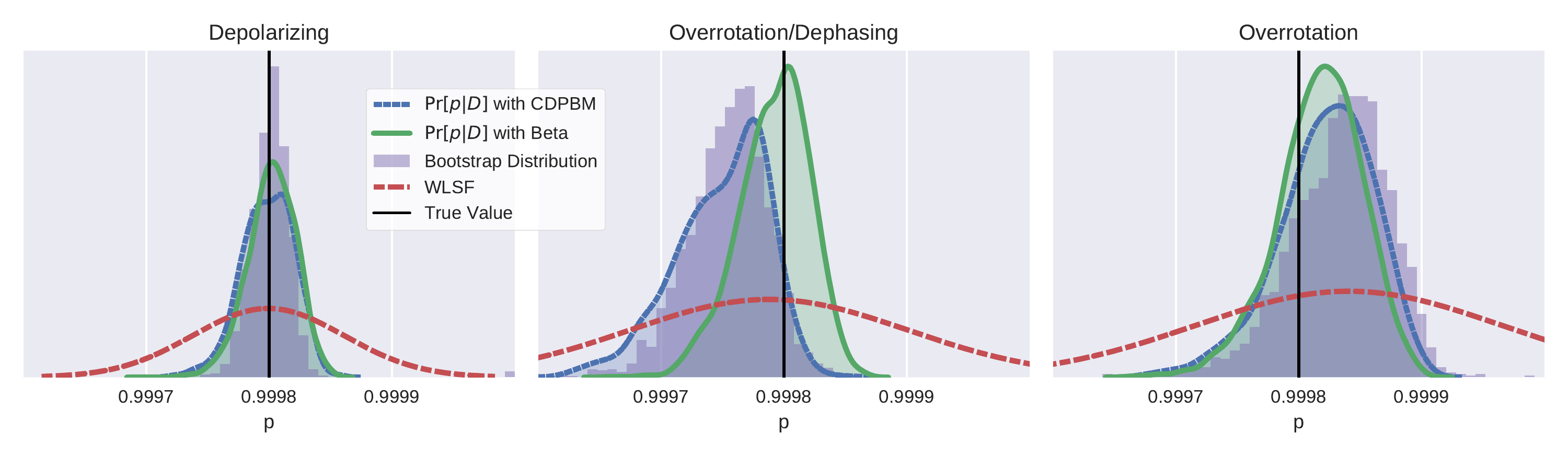}
    \caption{
    For each of the three noise models defined in 
    \autoref{sec:rb-various-noise-models}, four types of data
    processing are performed to compare their estimates of 
    (and uncertainties in) the parameter $p$ from the standard 
    RB protocol.
    Each dataset consists of $I=20$ random sequences with
    $N=30$ repetitions each at each of the sequence lengths 
    $1$, $100$, $200$, $500$, $1000$, $2000$, $5000$, 
    $10000$, $20000$ and $50000$.
    The first two methods show the posterior marginal of $p$ under 
    the models from \autoref{eq:dp-first-moment} and 
    \autoref{eq:betabin-model}, respectively.
    The next two methods are non-parametric 
    bootstrapping and weighted least squares
    fitting, as described in \autoref{sec:frequentist-approach}.
    }
    \label{fig:different-noise-models-comparison}
\end{figure*}

\begin{figure*}
    \includegraphics[width=\textwidth]{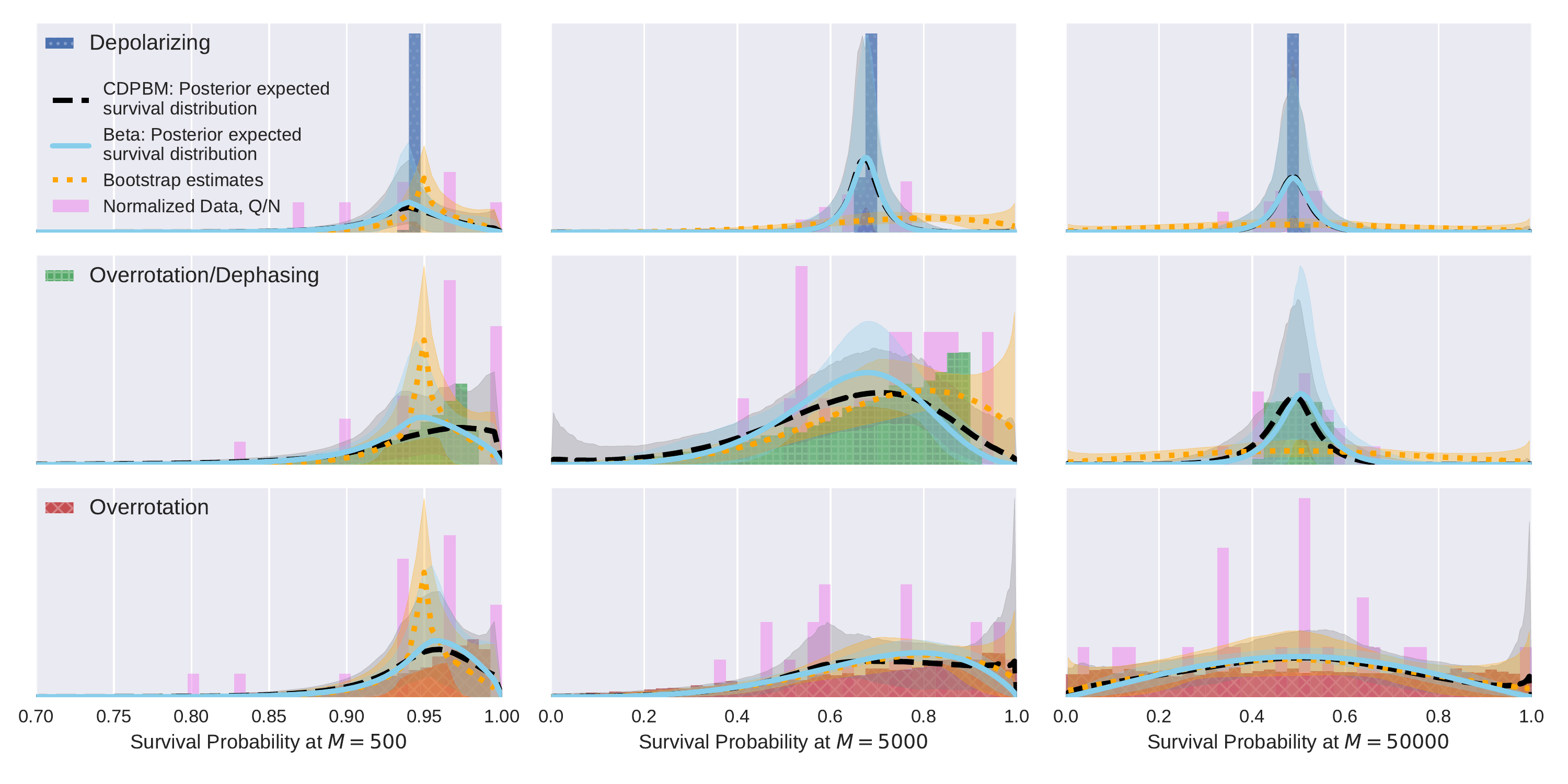}
    \caption{
    Although the survival distributions at each sequence length 
    are considered to be nuisance parameters of the model, their
    posteriors are nonetheless interesting and provide a diagnostic 
    check.
    Here, the three rows correspond to the noise models described in
    \autoref{sec:rb-various-noise-models}, and each column is a 
    different sequence length.    
    In each plot, Bayes' estimate of the survival distribution is shown 
    for both models \autoref{eq:dp-first-moment} and 
    \autoref{eq:betabin-model} along with their pointwise 
    95\% credible envelopes.
    Similar 95\% confidence envelopes are shown for the bootstrap method.
    These are overlaid on top of histograms sampled from the true 
    survival distributions, as well as the (normalized) data 
    that were actually used in the inference.
    }
    \label{fig:different-noise-models-survival-dists}
\end{figure*}

This dataset was processed in a few different ways.
Posterior results using the CDPBM-survival-distribution 
model \autoref{eq:dp-first-moment} 
are summarized in \autoref{fig:different-noise-models-summary}.
The slightly simpler Beta-survival-distribution 
model \autoref{eq:betabin-model} was also used, which is compared 
to the CDPBM model in \autoref{fig:different-noise-models-comparison},
along with weighted least squares fitting, and
a non-parametric bootstrap with 2000 samples.
Additionally, estimates of the shapes of some survival distributions
are seen in \autoref{fig:different-noise-models-survival-dists}.
The prior distribution on the tying parameters was chosen to be
$\pi(p,A,B)=\uniform([0,1]^3)$ in all cases.

\subsection{Low Data Regime}
\label{sec:low-data-regime}

One advantage of using the full likelihood model is that it transitions 
seamlessly to low data regimes where normal approximations fail and 
the usual sample moments are ill-defined.
At a given sequence length, if we only pick a handful of sequences $I$
with a handful of shots $N$ each, then there is a good chance
that $Q_{M,e,i}$ will be equal for all $i=1,...,I$.
This is especially true near the boundaries 0 and 1.
In this event, it is difficult to use a 
weighted least-squares fit.

To illustrate our Bayesian model in this regime, we consider simulated data 
from standard RB using the gate dependent overrotation model from 
\autoref{eq:overrotation-model}.
We choose this model because it has very wide survival distributions,
as seen in \autoref{fig:different-noise-models-survival-dists}.

We wish to demonstrate that posterior distributions in the low-data
regime meaningfully report the parameter of interest, $p$.
The worst thing an inference method can do in this example is predict
that the RB parameter $p$ is larger than it actually is. 
Therefore instead of summarizing a posterior in terms of its mean value
(Bayes' estimate), it is more helpful to summarize it in terms of 
the the value at a one sided credibility level $\alpha$,
\begin{align}
    p_\alpha(D) &= \left[p_0\text{ such that }\Pr(p > p_0|D)=\alpha\right].
\end{align}
Here, $\Pr(p|D)$ is the posterior of $p$ under the beta model 
\autoref{eq:betabin-model} with the 
same prior as in \autoref{sec:rb-various-noise-models} given the RB dataset $D$.
For example, according a given posterior, 
with 95\% probability, $p_{0.95}(D)$ should be a lower 
bound for the true the value of $p$.
Fixing the model and the prior, the quantity $p_\alpha(D)$ is itself
a random variable as it depends on $D$.
What we desire in our numerical test is that consistency condition
\begin{align}
    \Pr(p_\alpha(D) < p_\text{true}) &\geq \alpha
\end{align}
is satisfied for any level $\alpha$ that we care about.

\begin{figure*}[!ht]
    \includegraphics[width=\textwidth]{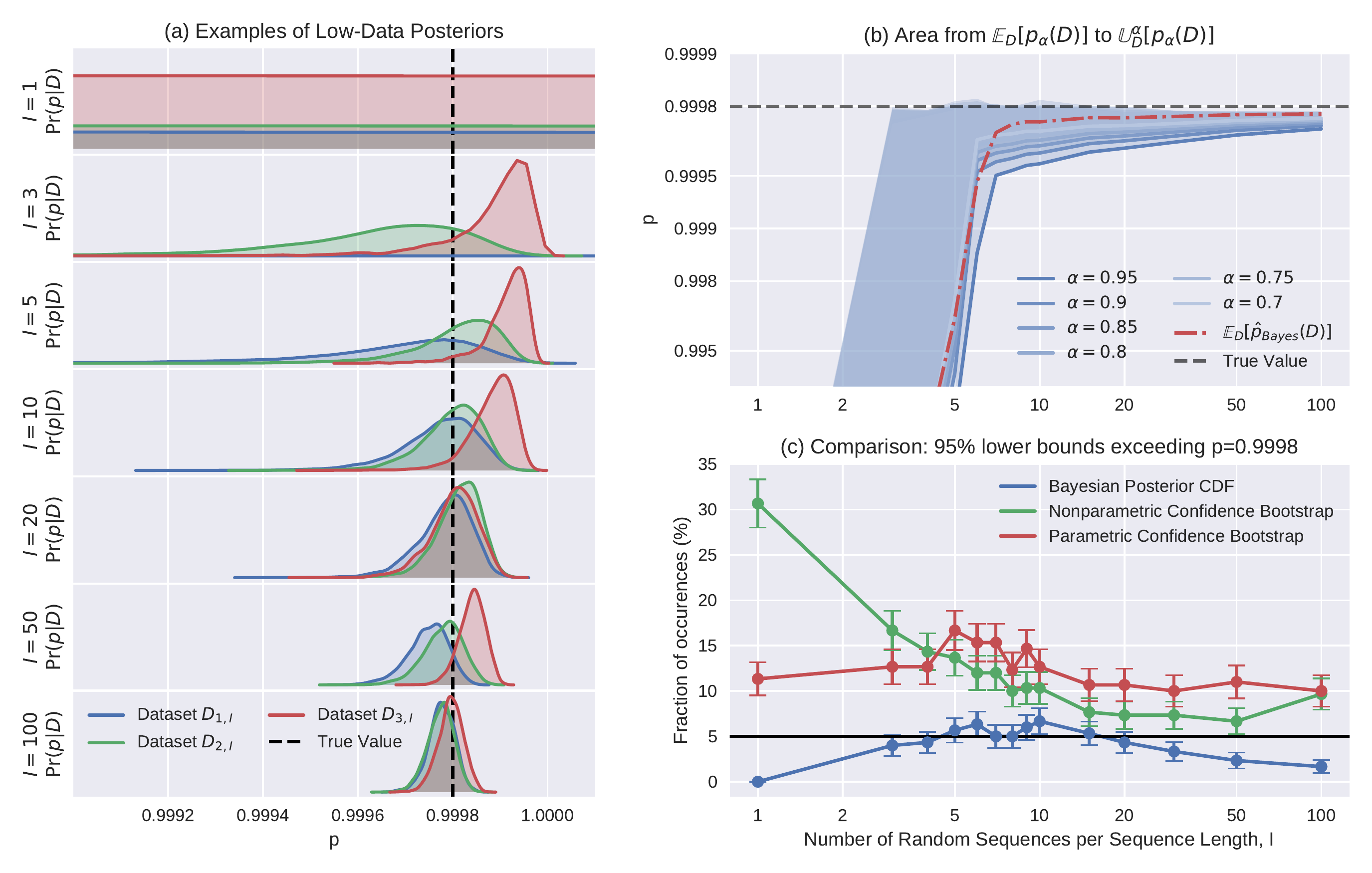}
    \caption{
    Data from the overrotation model \autoref{eq:overrotation-model} 
    was simulated 300 times for several values of $I$, the number of
    random sequences per sequence length. 
    In all cases $|\M|=10$ sequences were used 
    with $N=5$ repetitions of each random sequence.
    Posteriors were computed for every dataset, $p$-marginals
    for three of which are shown in (a) for several values of $I$.
    (b) The area between the upper $(100\cdot\alpha)$\% quantile and the mean
    value of $p_\alpha(D)$ is shown for several values of $\alpha$,
    demonstrating the posterior's ability to reliably report credible
    lower bounds for $p$.
    (The average value of Bayes' estimate is shown for comparison.)
    (c) Finally, we isolate the $\alpha=0.95$ case and display it
    along with bootstrapped lower 95\% confidence bounds, which do
    not stay under the desired line.
    These fractions were computed by running the bootstrap method on the 
    same collections of 300 data sets.
    Error bars are single standard deviations of simple binomial statistics.
    }
    \label{fig:low-data}
\end{figure*}

To evaluate this criterion we compute $p_\alpha(D)$ for many 
simulated datasets $D$.
Each dataset uses the sequence lengths
\begin{align*}
    \M=\{1,100,200,500,1000,2000,\\\quad\quad5000,10000,20000,50000\}
\end{align*}
and the repetition number $N=5$.
Three-hundred data sets were considered at each of the values
$I=1,3,5,10,20,30,50,80,100$.
\autoref{fig:low-data} shows both a selection of posteriors, as 
well as a summary of the distribution of $p_{0.95}(D)$ at 
each value of $I$.
Note that the sharp elbow displayed in \autoref{fig:low-data}(b) 
could be used in practice to decide on an appropriate amount of data 
to take: in this example, there is a huge advantage in moving
from $I=5$ to $I=10$, but not much of an advantage in moving
from $I=10$ to $I=15$.

The bootstrapped confidence bounds
discussed in \autoref{sec:frequentist-approach} 
are also sensibly defined in the low data regime.
In \autoref{fig:low-data}(d), however, we see in both
parametric and non-parametric bootstrapping that the
MLE has a tendency to exaggerate confidence.
All bootstrap distributions contain 600 samples.

\subsection{A pathological model: pushing the Dirichlet process to its limits}

\begin{figure}
    \includegraphics[width=0.6\textwidth]{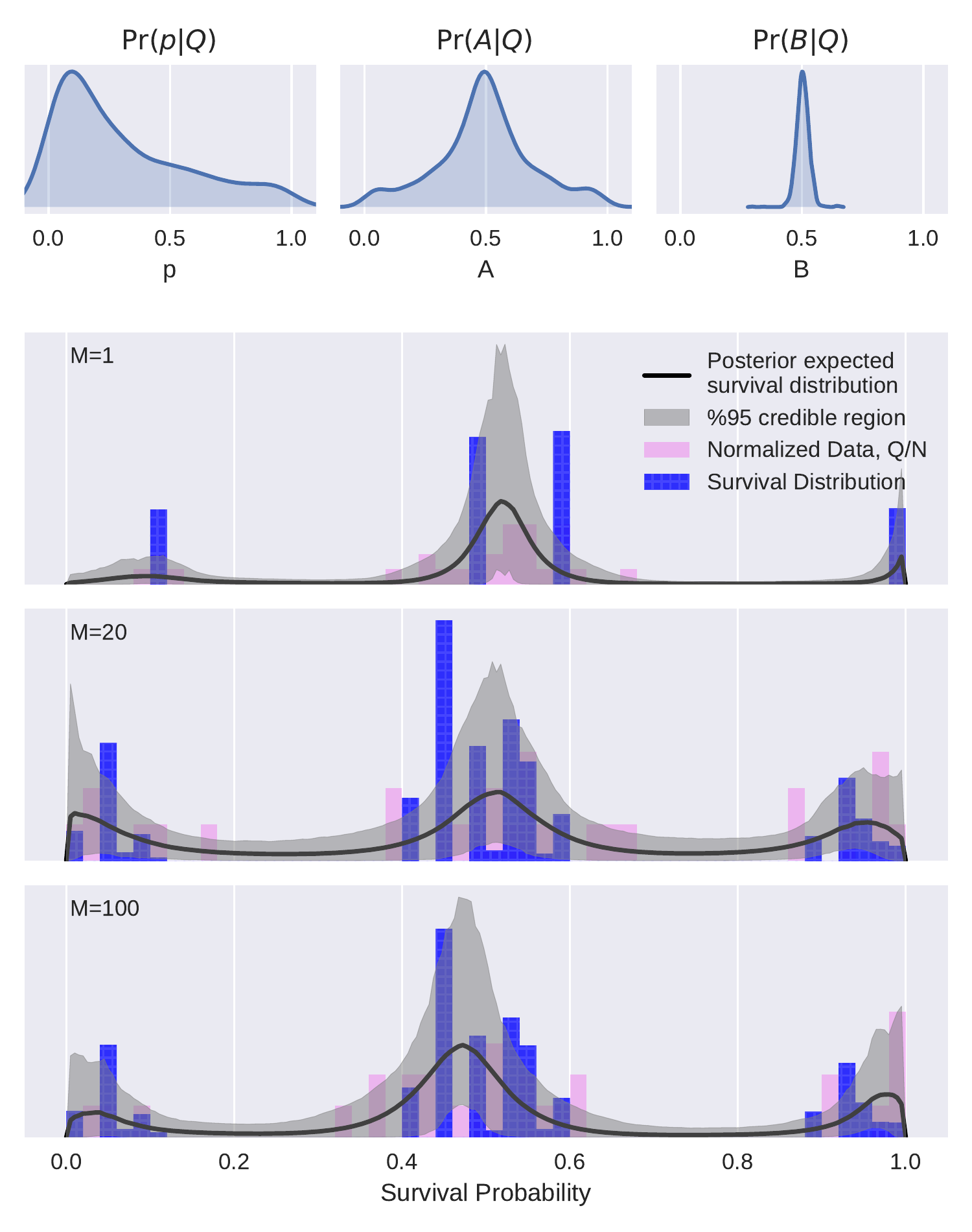}
    \caption{
    The top row of three plots show marginal posterior distributions 
    of the standard RB protocol tying parameters with data 
    simulated according to the 
    pathological noise model defined in \autoref{eq:pathological-error}.
    The bottom column of three plots show posterior summaries of the
    survival distribution at the sequence lengths $M=1,20$ and $100$,
    respectively.
    }
    \label{fig:pathological-distributions}
\end{figure}

To demonstrate that CDPBM based models are capable of handling strange 
underlying survival distributions, we use a highly
pathological error model, constructed to have multiple distinct peaks.
The model has gate-independent qubit noise defined as the convex mixture 
of a channel that resets to a fixed pure state, a channel that resets to
identity, and the identity channel, or explicitly
\begin{subequations}
    \begin{align}
        \E(\rho)
            &= \Tr(\rho)\left(
                p_1 \ketbra{\psi_r} + p_2\frac{\I}{2}
            \right) + (1-p_1-p_2)\rho.
    \end{align}
    \label{eq:pathological-error}
\end{subequations}
We used the parameters $p_1=0.9$, $p_2=0.001$, and 
$\ket{\psi_r}=\ee^{-\ii 0.05 (X+Y)}\ket{0}$ in our simulations.
This noise model results in an average gate fidelity of 
$0.5495$, or a decay base of $p=0.099$.
Due to the high value of $p_1$, this error channel is so bad that running RB 
as a characterization tool is not a great choice in the first place, 
and therefore looking at the posterior distribution
of $(p,A,B)$ is of little direct use, although similarly bad channels can arise when using interleaved RB to extract tomographic information \cite{kdr+_robust_2014}.
In any case, we provide certain marginals
at the top of \autoref{fig:pathological-distributions} anyway.
However, our point is to look at the posterior of the survival parameters,
$\vec{x}_S$, which are summarized in the bottom section of \autoref{fig:pathological-distributions}.
This posterior was computed using the sequence lengths
$\M=\{1,2,5,20,50,100\}$ with $I=30$ random sequences per
sequence length, and $N=50$ repetitions each.
The same gateset as \autoref{sec:rb-various-noise-models} was used,
with the same initial state and measurement operators.

\begin{figure}[h]
    \includegraphics[width=0.8\textwidth]{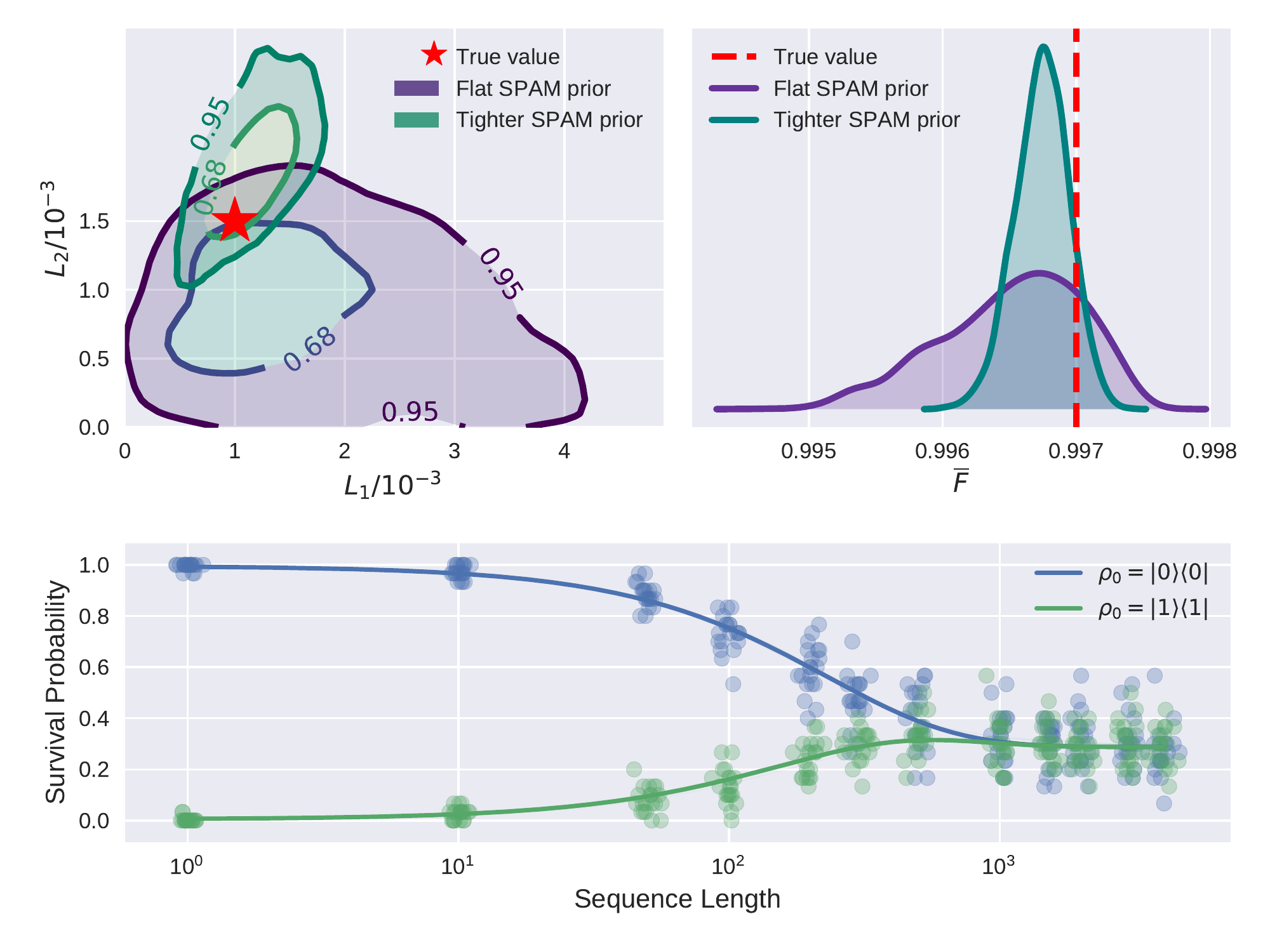}
    \caption{
    Posterior summaries for the LRB protocol under the model from
    \autoref{eq:betabin-model} and two different prior distributions.
    Simulated data was sampled at $|\M|=12$ 
    sequence lengths, each with 
    $I=15$ random sequences and $N=30$ repetitions per sequence.
    The joint posterior marginals of the 
    leakage and seepage parameters is shown (top left), as well as the 
    posterior marginals of the average gate fidelity (top right).
    The LRB tying functions are plotted using parameters randomly 
    drawn from the posterior tying distribution (bottom). 
    Superimposed are the normalized data, where each dot comes from 
    a unique random sequence.
    }
    \label{fig:lrb}
\end{figure}

\subsection{Complicated Tying Function: Leakage RB (LRB)}
\label{sec:lrb}

There are a few protocols which measure leakage
of information into and/or out of the qubit subspace
~\cite{wood_quantification_2017,wallman_robust_2015,chen_measuring_2016,chasseur_complete_2015}.
Here we provide an example using our framework with the LRB protocol that 
is described in Ref.~\cite{wood_quantification_2017} with an experimental 
implementation reported as a part of Ref.~\cite{mckay_efficient_2016}.
We have chosen this protocol because it has one of the most complicated 
tying functions of existing protocols; 
for a single qubit there are at least seven tying parameters, 
three of which are not nuisances.
Moreover, it is not quite a SPAM-free protocol---some
of the information that is necessary to decouple the three 
parameters of interest from each other
is contained in the constant offset term 
as well as the coefficients of the exponential terms.

We consider a system with a Hilbert space $\X=\X_1 \oplus \X_2$, 
where $\dim\X_1=d_1=2$ and $\dim\X_2=d_2=1$, 
and $\X_1$ is the computational subspace.
Our noise model is gate independent, 
equal to the depolarizing leakage extension
(DLE)~\cite{wood_quantification_2017} of $\E_\text{dephasing}\circ\E_\text{rot}$
where
\begin{subequations}
\begin{align}
    \E_{\text{dephasing}}(\rho) &= (1-s)\rho+sZ \rho Z \\
    \E_\text{rot}(\rho) &= \ee^{-\ii \alpha Z/2}\rho\ee^{\ii \alpha Z/2},
\end{align} 
\end{subequations}
and where we denote the resulting DLE as $\E$.
The parameters $L_1$ and $L_2$ are called the leakage and seepage 
respectively, and are given by
\begin{subequations}
\begin{align}
    L_1 &= 1-\Tr\I_1\E(\I_1/d_1) \\
    L_2 &= \Tr\I_1\E(\I_2/d_2)
\end{align}
\end{subequations}
where $\I_1$ and $\I_2$ are the projectors onto $\X_1$ and $\X_2$.
We see that the leakage quantifies how much population from $\X_1$ leaks 
out of $\X_1$, and the seepage quantifies how much population seeps into 
$\X_1$ from $\X_2$.
We have assumed that our initial states are prepared in $\X_1$
for simplicity in this demonstration.
We use the values $s=0.003$, $\alpha=0.1^\circ$, $L_1=0.001$, and $L_2=0.0015$.
The average gate fidelity of $\E$ averaged over states in 
$\X_1$ comes out as $\overline{F}=0.997001$ with these numbers.

One feature of the fitting method proposed along with the LRB
protocol is that it implicitly
asserts that certain SPAM parameters sum to unity, and 
certain other SPAM parameters sum to zero (respectively $A$
and $B$ in our appendix).
Though this may be valid for some systems, it depends on the methods of 
state preparation and measurement for the given device.
We have highlighted our ability to loosen this assertion by 
comparing the posterior distributions due to two priors.
In the first, all SPAM parameters have flat non-informative priors,
and in the second, prior information is introduced that causes the two sums
in question to have support of roughly $\sim1(\pm 0.05)$ 
and $\lesssim 0.05$, respectively.
Explicit details of this prior, along with the LRB protocol and how we slightly modified its parameterization can be found in \autoref{app:reparam-leakage-rb}.
Posterior results are summarized in \autoref{fig:lrb}.

\section{Departing from Bernoulli Trials}
\label{sec:departing-from-bernoulli-trials}

All models thus far have assumed that measurements conditional on some 
sequence length and experiment type are Bernoulli trials, or stated 
differently, we have assumed that two-outcome strong measurements are performed.
For some quantum systems, this is not possible, with some other non-binary
result being returned from a measurement operation.
It would be nice to be able to analyze data from  
RB and related protocols for these systems too.
In this section we point out that our methods extend straight-forwardly
(at least in principle) to other measurement schemes.

For example, we can extend the model from \autoref{eq:dp-first-moment} 
to the case of referenced photon counts from a Nitrogen Vacancy
center in diamond.
In the most commonly used measurement scheme for this system, instead of 
having direct access to Bernoulli trials with the probability 
$q=\Tr \tilde{E}_e \tilde{\G}_{\vec{j}}(\tilde\rho_e)$ 
for some sequence $\vec{j}$,
we instead have obstructed access to this quantity through
the random triplet $(X,Y,Z)|\alpha,\beta$ where $0<\alpha<\beta$ 
are unknown Poisson rates \cite{hincks_statistical_2018}, giving
rise to the likelihood
\begin{subequations}
	\begin{align}
		q|G 
		    &\sim G \\
		(X,Y)|\alpha,\beta
		    &\sim \poisson(\alpha)\times \poisson(\beta) \\
		Z|q,\alpha,\beta
		    &\sim \poisson(\beta
		        +(\alpha-\beta)q)
	\end{align}
	\label{eq:nv-dp}
\end{subequations}
where the prior is exactly the same as in \autoref{eq:betabin-model} 
or \autoref{eq:dp-first-moment}.
The subscripts ${M,e,i}$ were dropped for the sake of brevity.

\section{Conclusions and Outlook}
\label{sec:conclusions}

We have presented a Bayesian approach to analyzing data from RB+ experiments.
We used a formal framework to describe such protocols to emphasize that
RB and its derivative protocols, from the perspective of statistical inference,
are all quite similar.
Specifically, they all admit noise model dependent survival distributions
which are tied together parametrically by a combination of quantities of 
interest and nuisance (SPAM) parameters.
A handful of examples are summarized in \autoref{tab:models}.

We proposed a hierarchical Bayesian model that was 
constructed to be agnostic to the 
nature of these survival distributions, and hence to the noise model.
This was achieved by modeling them non-parametrically through Dirichlet
process priors.
We also considered modeling them parametrically through the Beta
distribution family.
For physically reasonable noise models we found that this simpler family 
worked well.
Therefore we suggest using the non-parametric model in, for example, first
runs where the system is not well understood, 
possibly switching to the parametric model when the system is 
better characterized and RB+ is being used for tune-ups.

Under either model, however, one ends up with a marginal posterior distribution of the
RB+ parameters, from which figures of merit can be computed.
We found qualitative similarity between the nonparametric MLE bootstrap distribution and the posterior distribution of the 
Bayesian nonparametric model when using a diffuse prior,
which merits further study.

We tested our Bayesian models under various noise types, data regimes, and 
protocols.
Our posterior distributions were computed numerically by drawing
posterior samples with MCMC methods.
As well as fitting well to survival distributions from standard error
models (\autoref{fig:different-noise-models-summary}), we were also able to fit to pathological multi-modal survival
distributions (\autoref{fig:pathological-distributions}).
Due to our choice of parameterization, estimating probabilities
very close to the boundaries $[0,1]$ is stable.
Of particular importance, we found no systematic tendency to over-report 
gate qualities.
Specifically, a numerical study of standard RB in the 
low data regime showed that posteriors of our model accurately report 
uncertainty---for example,
a 95\% credible lower bound on the fidelity is indeed a 
lower bound to the true value at least 95\% of the time (\autoref{fig:low-data}).
This is in contrast to the frequentist bootstrapping techniques
we compared to, which do not always pass this sanity test in the low-data
regime.

We assumed throughout this work that the model being used for a given dataset 
was correct.
In practice, features like non-Markovian noise may necessitate corrections
to a model.
A useful direction of research would therefore be to explore Bayesian 
model selection and cross validation.

\acknowledgements{IH thanks Robin Blume-Kohout for helpful
correspondence regarding frequentist estimators for RB, 
and Thomas Alexander for numerical MCMC advice.

IH, JJW, and DGC gratefully acknowledge contributions from the Canada First 
Research Excellence Fund, Industry Canada, Canadian Excellence
Research Chairs, the Natural Sciences and Engineering
Research Council of Canada, the Canadian Institute for Advanced
Research, and the Province of Ontario. CF was supported by the Australian Research Council
Grant No. DE170100421.
CG was partially supported by the Australian Research Council (ARC) via the Centre of Excellence in Engineered Quantum Systems (EQuS) project number CE110001013 and by the US Army Research Office grant numbers W911NF-14-1-0098 and W911NF-14-1-0103.
}


\bibliographystyle{apsrev4-1}
\bibliography{brb}


\appendix

\section{Sampling Strategies}
\label{app:sampling-strategies}

For the complete details of our numerical methods there is no better place
to look than the code base that accompanies this paper\cite{code_repo}.
In this section, we summarize---at a high level---some of the tools and 
tricks we used.

\subsection{Posterior Sampler}

Analytic formulae for the posteriors of our models are intractable---we 
must instead choose a numerical inference algorithm to sample points 
from the posterior.
A sufficient number of these points can be used to compute any quantity
of interest related to the posterior.
We used the Hybrid Monte Carlo (HMC) sampler using the No-U-Turns (NUTS) 
heuristic \cite{duane_hybrid_1987,hoffman_no-u-turn_2014}.
This is a Markov chain Monte Carlo (MCMC) sampling strategy 
which has gained widespread use due to its lack of tuning parameters, 
fast mixing rate, and ability to handle large
numbers of parameters.
We provide a very brief introduction to MCMC algorithms in
\autoref{app:mcmc-intro}.
Specifically, we used the \pystan~interface to the \stan~
library \cite{carpenter_stan:_2017}.
Probabilistic programs such as those written 
in \autoref{eq:betabin-model} and \autoref{eq:dp-first-moment} 
can entered nearly verbatim as input to this library (or other 
similar libraries), and samples from the posterior are returned.
We suspect that sampling algorithms customized to our models could
significantly outperform these generic tools, but it is hard to turn down the 
convenience of modern probabilistic programming languages and 
automatic differentiation.

It warrants mention why we have not used sequential Monte Carlo (SMC),
which has emerged as a popular inference engine for quantum information
processing tasks~\cite{granade_qinfer:_2017}, including
for RB and IRB with $N=1$~\cite{granade_accelerated_2015}.
Our main reason is that we wish to leave open the 
option of sampling from exact posterior distributions, especially 
while still in the proof-of-principle stage.
SMC operates by storing the distribution over parameters as 
a weighted mixture of delta functions.
Data is entered sequentially and the prior is gradually transformed into
the posterior with the inclusion of each subsequent individual datum.
While SMC uses exact likelihood functions to sequentially update the
weights with Bayes' rule, it also occasionally requires a resampling 
operation that moves the positions of the delta functions to where 
they are most needed.
This resampling step usually only considers the first two moments
of the distribution, and tends to distort the distribution toward being
multivariate normal---see Appendix B of
Reference~\cite{granade_qinfer:_2017}.
Therefore, in SMC, posterior distributions are convolved with normal approximations to
the true posterior distribution.
However, SMC has an important advantage in that it can naturally 
be used with adaptive experiments, where the next experiment (sequence length,
measurement type, etc.) is chosen based on the current state 
of knowledge.
Also, SMC is often less computationally expensive and always 
highly parallelizable.

\subsection{Reparameterizations}

MCMC samplers benefit from using an optimal parameterization of the model ---
simply reparameterizing a model can make huge differences to the convergence, mixing rates, and stability.
Ideally, posterior parameters are decorrelated, centered at 
the origin, and have a variance of order unity.
Doing a perfect job at this would require knowing the posterior in advance
of sampling from it, so we must instead rely on other heuristics.

For example, samplers have trouble near hard cutoffs, requiring
special boundary specifications, and time can be wasted
proposing random-walk values outside of the allowed region.
This is relevant to our models where it is common to be inferring
values that are physically restricted to the interval $[0,1]$, and that are 
ideally very close to the boundary, such as the average gate fidelity of 
a gate-set.
Modern Bayesian inference libraries, such as \stan, will 
automatically remove hard cutoffs
by reparameterizing the model through a logit function for interval constraints,
or through the logarithm for one-sided constraints.
(It also multiplies the pdf by the change of variables Jacobian 
so that the prior is not distorted.)

We can do slightly better than this if we have prior expectations about
some parameter values.
For instance, if we expect a decay parameter $p$ to be on the order of 
$p_0=0.9999$, then instead of using the sampling variable 
$\tilde{p}=\logit(p)$ as would automatically be done by \stan,
we can use the variable $p_\err$ where $p=\logit(p_0+p_\err)$.
This is distinct from the role of a prior distribution in the sense that the correct distribution is sampled from even if we have set $p_0$ far from the value $p$ that we are attempting to estimate; rather, we arrive at our samples less efficiently in that case.
For reference, $\logit(0.9999)\approx 9$, and we will have, with very
little effort, prevented the sampler from making an initial random walk
from $0$ to $9$ while also keeping track of $200$ other variables.

If we additionally have expectations about the standard deviation, 
say we expect $\delta p=p_0\pm \delta p$ in the posterior,
then we can use the changed variable $p_\err$ where 
$p=\logit(p_0+\delta \tilde{p}\cdot p_\err)$.
If we let $\delta\tilde{p}=\delta p \cdot p_0^{-1}(1-p_0)^{-1}$, then 
$p_\err =0\pm 1$ will translate to $p=p_0\pm \delta p$.
This trick does not affect the posterior in any way, it 
only improves sampling performance.
We have had success using least-squared fits to estimate $p_0$ and/or 
$\delta p$
(along with other parameters), using these values in the parameter
transformation.
If there is not enough data to meaningfully estimate $\delta p$ with, for
example, a weighted least squares fit, then $\delta\tilde{p}=0.5$ is a
fine choice.

The above heuristic should apply well to most probability parameters.
There is a notable exception that comes up in low data regimes,
which we will now illustrate in the case of standard RB for concreteness.
Here, the tying function is $(A-B)p^M +B$, and
for high quality devices, and at very low values of $M$, the survival 
probability is roughly equal to $A\lesssim 1$.
Moreover, low values of $M$ are exactly where we learn the most about 
$A$, allowing us to decorrelate its value from $p$ and $B$.
Suppose, however, that we are in the low data regime defined 
by
$1/(1-A) \gg N\cdot I$, so that at the lowest values of 
$M$ it's very likely that every single shot of the experiment will return $1$.
In this case any estimation technique will only be capable of producing 
a lower bound on the value of $A$; any value of $A$ arbitrarily close to 
$1$ will be consistent with the data.
This is a problem for the $\logit$ rescaling discussed above because 
an estimate of $A$ arbitrarily close to $1$ implies a sampling 
parameter $A_\err$ that is arbitrarily large no matter the choices 
of $A_0$ and $\delta A$.
There are a few potential paths forward.
One is to switch sampling strategies to something like Riemannian
Manifold HMC that fairs better with varying curvature in
parameter space~\cite{betancourt_generalizing_2013}.
Another is to reparameterize in a different way, for example through
an exponential distribution.
Perhaps the easiest, however, is to recall that the lowest values of 
$M$ are dubious in the case of gate dependent noise, and no data
should be taken there anyway.
We can just take a new definition of the initial state to be our old
initial state acted on by a fixed number of random gates, effectively
lowering the value of $A$.

\section{Nonparametric Families}
\label{app:nonparametric-families}

\subsection{Dirichlet Processes}
\label{app:dirichlet-process}

Dirichlet processes (DP) can be introduced in many ways.
Given how they are used in the main body of this paper, 
we will introduce them as a natural extension to beta and Dirichlet priors
as follows.
Much more comprehensive introductions can be found elsewhere, for example,
see this article of Teh~\cite{teh_dirichlet_2011}.

First, consider of coin with an unknown bias $p$ that we wish to infer.
If we flip it $N$ times and sum the resulting number of heads 
we get the random variable $X\sim\Binom{N}{p}$.
In a Bayesian setting, we start by assigning a prior $\pi(p)$ to 
the unknown quantity $p$.
Having collected the variate $x$ of $X$, our posterior is 
proportional to $\Pr(p|x)\propto\int \binom{N}{x}p^x(1-p)^{N-x}d\pi_p(p)$.
An important property the beta distribution is that when 
it is used as the prior in this example, 
say $\pi(p)=\betadist(a,b)$ for some choices $a>0$ and $b>0$,
then this integral has a nice closed form solution,
\begin{align}
    \Pr(p|x) = \betadist(a+x,b+N-x).
\end{align}
This is one of the reasons the beta distribution family is the canonical 
family of distributions with support on $[0,1]$.
Moreover, from this formula, an operational interpretation 
of the prior parameters
$a$ and $b$ is apparent: $a$ can be thought of as the number of prior
`heads' observations, and $b$ as the number of prior `tails' observations.
For example, $\pi(p)=\betadist(1,1)$ is asserting that one's prior knowledge
of $p$ is equivalent to having already flipped the coin twice, with each 
a heads and a tails landing once.

Let us generalize one step further before mentioning Dirichlet processes.
Suppose we are interested in inferring the weights $p=(p_1,...,p_K)$ 
of a $K$-sided die, where $p$ is a finite probability distribution, so that
$\sum_k p_k=1$ and $p_k\geq 0$.
(The coin example above is the case $K=2$.)
Rolling this die $N$ times and binning the number of times each side 
lands face up results in the random variable
$X=(X_1,...,X_K)\sim\operatorname{Multinomial}(N,p)$.
The Dirichlet distribution family is the natural extension to the beta 
distribution family for $K>2$.
Namely, if we set the prior $\pi(p)=\dirichlet(a_1,...,a_K)$, then 
the posterior distribution is given by
\begin{align}
    \Pr(p|x)=\dirichlet(a_1+x_1,...,a_K+x_K)
\end{align}
where $x=(x_1,...,x_K)$ is the data.
As before, this provides an operational interpretation of the prior parameters 
$a_1,...,a_K$ --- the value $a_k$ can be interpreted as the number of 
prior observations of side $k$ out a total of $\sum_k a_k$ prior observations.

Dirichlet processes can be thought of as the next logical step in this progression.
We move from probability distributions with two sides, to $K$ sides, and
now to a continuum of sides; 
Dirichlet processes are natural priors for 
probability density functions.
Suppose that $f$ is an unknown probability density function on the 
sample space $\Omega$ that we wish to infer.
Therefore $\int_\Omega f(x) d\mu(x)=1$ where $\mu$ is some measure on
$\Omega$.
Data is collected from this unknown distribution through the random 
variable $X\sim f$.
We wish to set our prior on $f$ to be a Dirichlet process, which is 
a distribution of distributions on the sample space $\Omega$.
First, we need to define what a Dirichlet process is: given a
distribution $G_0$ defined on $\Omega$ and a positive real number $\alpha>0$,
we say that the random distribution $G$ is Dirichlet process distributed with base distribution $G_0$ and concentration parameter $\alpha$,
writing $G\sim\dirp(\alpha,G_0)$ to denote this, if for any finite disjoint
measurable partition $\cup_{k=1}^K B_k =\Omega$, it holds that
\begin{align}
    (G(B_1),...,G(B_K))\sim \dirichlet(\alpha G_0(B_1),...,\alpha G_0(B_K)).
\end{align}
Note that for $B\subset \Omega$, all we mean by $G(B)$ is the 
probability of an event in $B$ under distribution $G$.
This means that $\alpha G_0(B)$ has the interpretation of being the 
number of prior observations in the region $B\subset \Omega$, and 
that to be Dirichlet process distributed means to be a distribution 
which obeys this
condition for every possible partition of $\Omega$ into regions.
In our previous example with the $K$-sided die,
we could have reparameterized the 
Dirichlet prior as $\dirichlet(\alpha,g)$ where $\alpha:=\sum_k\alpha_k$ and 
$g:=(\alpha_1/\alpha,...,\alpha_K/\alpha)$ to be more
notationally analogous to the present example.

If we let $\pi(f)=\dirp(\alpha, G_0)$ be the prior 
distribution of $f$, and suppose we make $N$ iid measurements
$X\sim f$,
then the posterior is also Dirichlet process distributed, with
\begin{align}
    \Pr(f|x)
        &=  \dirp\left(
		\alpha+N, \frac{\alpha}{\alpha+N}G_0
		+ \frac{N}{\alpha+N}\frac{\sum_{k=1}^N \delta_{x_k}}{N}
	\right)
\end{align}
where $\delta_{x_k}$ is the delta distribution centered at 
the datum $x_k\in\Omega$.
We see that the base distribution of the posterior of $f$ is a 
mixture of the prior's base distribution and the empirical
distribution of the data.
It also makes it clear that $\alpha$ still has the interpretation as the
total number of prior observations.
Despite these nice interpretations, so far it might seem like 
Dirichlet processes are too abstractly defined to make them 
practical; at the end of this section, we will see that they have a
alternate and procedural description which is not too hard to work with.

Dirichlet processes can be used as a generic stand-in for 
parametric priors in Bayesian models.
For example, suppose we have samples $X_k|\lambda\sim \poisson(T_k\lambda)$
for some rate of events $\lambda$ measured for durations
of time $T_k$.
Moreover, suppose that there is not just one underlying 
rate of emission, but that there truly is a distribution of rates 
taking place, and we would like to infer what this distribution looks like.
A parametric Bayesian approach might be through the model
\begin{align}
	X_k|\lambda &\sim \poisson(T_k\lambda) \nonumber \\
	\lambda &\sim \gammadist(a,b) \nonumber \\
	a,b &\sim \pi(a,b)
\end{align}
where $\pi(a,b)$ is some prior on $a$ and $b$.
We have parameterized the unknown distribution over rates with 
a gamma distribution with hyperparameters $a$ and $b$; if we infer
$a$ and $b$, we can plot an estimate of the distribution of $\lambda$.
Hovever, if we are unable to confidently assert that the distribution over
rates $\lambda$ must be gamma distributed, then we might choose a \textit{nonparametric} Bayesian approach with the model
\begin{align}
	X_k|\lambda &\sim \poisson(T_k\lambda) \nonumber \\
	\lambda &\sim G \nonumber \\
	G &\sim \dirp(\alpha,G_0).
	\label{eq:simple-nonparametric-model}
\end{align}
Just as we looked at the posterior of the parameters $a$ and $b$ above,
here we can look at the posterior of $G$.

One caveat to random distributions drawn from
$\dirp(\alpha,G_0)$ is that they are almost 
surely discrete in nature, even when $G_0$ is 
a continuous distribution.
With probability one, $G\sim\dirp(\alpha,G_0)$ will be
of the form
\begin{align}
	G(\cdot)=\sum_{k=1}^\infty p_k \delta_{\theta_k}(\cdot)
	\label{eq:discrete-dist}
\end{align}
where the $p_k$ are probabilities summing to unity and $\theta_k$ 
are members of $\Omega$.
However, this is not a big deal in practice for two reasons.
The first is that any continous function can be approximated with 
arbitrary accuracy in $L_1$ distance using distributions 
of the form \autoref{eq:discrete-dist}.
Secondly, we always have the option of convolving $G$ with 
some smooth distribution to end up with a smooth distribution.
This is called a Dirichlet process 
\textit{mixture model}~\cite{escobar_bayesian_1995}.

Sethuraman found a way to construct instances of $\dirp(\alpha,G_0)$ 
in the form of \autoref{eq:discrete-dist} using a 
\textit{stick breaking process} \cite{sethuraman_constructive_1994}.
A random variate $G\sim\dirp(\alpha,G_0)$ can be construction as 
follows.
The points $\theta_k$ are simply drawn from $G_0$ independently
and identically.
Their weights $p_k$, however, are derived from the following process.
A stick of unit length is broken in two at the random location
$V_1\sim\betadist(1,\alpha)$.
The first piece is kept and its length is assigned to the first 
weight, $p_1=V_1$.
The remaining piece has length $1-V_1$ and is broken again at a random 
fraction $V_2\sim\betadist(1,\alpha)$ of its length.
The first piece is kept and its length is assigned to the second 
weight, $p_2=V_2(1-V_1)$.
This process is repeated until the stick has been broken up
a countably infinite number of times, giving $p_k=V_k\prod_{l<k}(1-V_l)$.
We therefore have the representation
\begin{align}
	G &= \sum_{k=1}^\infty \left[V_k \prod_{l=1}^{k-1}(1-V_l)\right]\delta_{\theta_k} \nonumber \\
	V_k &\iid \betadist(1,\alpha),\quad
	\theta_k \iid G_0
	\label{eq:dp-stick-breaking}
\end{align}
which is equivalent to $G\sim\dirp(\alpha,G_0)$.

Finally, we remark that it is standard practice to assign 
a distribution to the parameter $\alpha$, acknowledging one doesn't 
know \textit{a priori} how good the base distribution $G_0$ is.
We can see this in the stick breaking process, where low values of
$\alpha$ lead to few important modes, and high values of $\alpha$ lead
to many modes.
In practice, Dirichlet processes are parameterized by their 
weights and locations, and the number of possible modes is truncated.
One can verify that a certain truncation is sufficient by making sure 
the last weights (which must decrease in size) 
are negligibly small.

\subsection{Constrained Dirichlet Process Beta Mixtures}
\label{app:cdpbm-motivation}

We wish to modify the Dirichlet Process, defined in the previous section,
so as to make it a suitable prior for survival distributions.
In order for such a prior to work well with state-of-the-art MCMC
samplers, which depend on gradients, we require a sample space 
of smooth distributions.
This is easily done, as in the previous section, by convolving variates
of the Dirichlet process with smooth distributions.
It is natural for us to convolve with Beta distributions rather
than the typically used normal distributions because
survival distributions have support only within the interval $[0,1]$.

The main difficulty of our construction lies in our second demand,
which is the ability to constrain certain moments of these random distributions
to specific values.
We draw inspiration from Yang et al.~\cite{yang_semiparametric_2010} 
who propose a method to specify 
either or both of the first two moments of Dirichlet process variates.
This method consists simply of shifting and scaling the delta locations 
$\theta_k$ (see \autoref{eq:dp-stick-breaking}) 
so that the mean and variance of $G$ are as desired.
We cannot use this approach directly because our domain is $[0,1]$;
for example,
we might need to shift some of our locations $\theta_k$ to be outside 
of this interval to obtain the correct mean, which is not allowed.
To overcome this, we use the logit function and its inverse
to constrain and unconstrain variables between $\Real$ and
$(0,1)$, as follows.

Using a sample space $\Omega=\Real\times (0,1)$, we 
begin by drawing a standard Dirichlet process distributed variate 
$\sum_{k=1}^{K}w_k \delta_{(\nu_k^*,r_k)}$.
Here, $r_k$ is the scaled variance parameter (\autoref{app:dist-reparam}) and 
$\nu_k^*$ represents an unconstrained beta mean.
We then constrain each of these latter values to $(0,1)$ by using the 
inverse logit function, $\nu_k=\logit^{-1}(\nu_k^*+h)=1/(1+\ee^{-\nu_k^*-h})$.
The value of $h$ is chosen as the unique real number which enforces the 
condition $\sum_{k=1}^K w_k \nu_k=\mu_1$.
This in turn guarantees that $\expect[G]=\mu_1$.
There is no analytic formula for $h$, but it can be found efficiently 
with numerical optimization.
In particular, Newton's method with an initial 
guess $h=\logit\left(\mu_1-\sum_{k=1}^K w_k \nu^*_k\right)$ has
quadratic convergence.
A code sample is shown in \autoref{lst:newton}.
This procedure produces a variate from what we call the 
mean-constrained Dirichlet process beta mixture mean-CDPBM distribution,
which is summarized in \autoref{eq:cdpbm} of the main body.

Draws from the second-moment-constrained version, 
$\CDPBM_K(\alpha, G_0, \mu_2)$, and the first-second-moment-constrained version,
$\CDPBM_K(\alpha, G_0, (\mu_1, \mu_2))$, are similar, except that 
a transform of the form $\logit^{-1}(h_1\nu^*+h_2)$ is necessary to 
constrain the variance as well as the mean.

If a protocol were to tie together the first two moments, following
\autoref{eq:dp-first-moment}, we would have the probabilistic program
\begin{widetext}
    \begin{subequations}
        \begin{align}
            \vec{x}_T 
                &\sim \pi(\vec{x}_T) \\
            \mu_{1,M,e}|\vec{x}_T 
                &= T(1,M,e,\vec{x}_T) \\
            \mu_{2,M,e}|\vec{x}_T 
                &= T(2,M,e,\vec{x}_T) \\
            \alpha_{M,e} 
                &\iid \gammadist(1,1) \\
            G_{M,e}|\alpha_{M,e},\mu_{1,M,e},\mu_{2,M,e} 
                &\ind \CDPBM_K\left(
                        \alpha_{M,e}, 
                        G_0,
                        (\mu_{1,M,e},\mu_{2,M,e})
                    \right) \\
            \mathclap{\rule{7.5cm}{0.2pt}} \nonumber \\
            q_{M,e,i}|G_{M,e} 
                &\ind G_{M,e} \\
            Q_{M,e,i}|q_{M,e,i} 
                &\ind \binomial(N, q_{m,i}).
        \end{align}
    \end{subequations}
\end{widetext}

\lstset{language=C++,
                basicstyle=\ttfamily,
                keywordstyle=\color{blue}\ttfamily,
                stringstyle=\color{red}\ttfamily,
                commentstyle=\color{green}\ttfamily,
                morecomment=[l][\color{magenta}]{\#}
}
\begin{lstlisting}[label=lst:newton,basicstyle=\small,caption={Stan 
        function (similar to C) 
        to transform input weighted locations ($\nu^*_k$) into output 
        locations ($\nu_k$) whose weighted mean is equal to $\mu$.}]
     
// nu_star is a vector of input locations
// w is a length-K vector of weights
// mu is the desired mean value   
vector compute_nu(vector nu_star, vector w, real mu) {


  real h;
  vector[size[w]] nu;
    
  // initial guess for h is exact when var(nu_star)=0
  h = logit(mu) - dot_product(w, nu_star);
    
  // fixed descent of five steps
  for (newton_loops in 1:5) {
    nu = inv_logit(h + nu_star);
    h = h - (dot_product(w, nu) - mu) / 
        (dot_product(w, nu .* (1-nu)));
  }

  nu = inv_logit(h + nu_star);
  
  return nu;
}
\end{lstlisting}

\section{MCMC Introduction}
\label{app:mcmc-intro}

A Markov chain Monte Carlo (MCMC) method is an algorithm
used to sample independent elements from some desired distribution using 
the following general principle: an instance of a Markov chain is 
simulated, where the Markov chain has been designed to have a steady-state
distribution equal to the distribution of interest.

The Metropolis--Hastings is one of the simplest such algorithms.
It is designed for the scenario where one wants to sample from 
the density function $f(x)$ but one only has access to an unnormalized 
version $g(x)$, where $f(x)=g(x)/\int g(x)dx$~\cite{hastings_monte_1970}.
This is often useful in the context of Bayesian inference 
where $f(x)=\Pr(x|d)$ is the posterior of $x$ given the data $d$.
By Bayes' law, $f(x)\propto\Pr(d|x)\pi(x)$ where both the likelihood
$\Pr(d|x)$ and prior $\pi(x)$ are known, but the normalization constant
$\int\Pr(d|x)\pi(x)dx$ is intractable.

The Metropolis--Hastings trick is to construct a Markov chain 
whose steady state 
distribution is given by $f(x)$, but for which 
simulating a random instance requires only evaluations of ratios
of $f$, which are the same as ratios of $g$.
Then we may start with an arbitrary initial value and evolve until we 
have reason to believe we are in the steady state,
which is determined either empirically or theoretically.
The last time sample represents a random sample drawn from $f(x)$.
If multiple samples are required, it is common to, say, throw out 
the first 1000 transient time points of the process (the
\textit{burn-in} period), and keep every 
100$^\text{th}$ subsequent time step as a random sample of $f(x)$.
A short auto-correlation time post burn-in, known as a fast \textit{mixing rate},
 is desired, so that fewer samples need to be thrown out.

The algorithm requires a \textit{proposal density} 
$h(x'|x)$ whose job is to propose the next value of the process, $x'$, given 
the previous value, $x$.
The prototypical choice is a normal distribution 
$h(x'|x)\propto e^{-(x-x')^2/2\sigma^2}$.
This choice affects the burn-in time and mixing rate.
For example, with a normal proposal density, a small variance will mean it 
takes many steps to move around the domain of $f$ leading to a slow
mixing rate.
On the other hand, a large variance may usually propose new locations
well outside the likely support of $f$, leading to high rejection rates 
and therefore also slow mixing.
A well-tuned proposal density will hit the sweet spot.

The algorithm is as follows:
\begin{enumerate}
	\item Somehow pick an initial value, $x_0$.
	\item For $k\geq 1$, draw a proposal and a random number,
		\begin{align*}
			x'&\sim h(x'|x_{k-1}) \\
			r&\sim\uniform([0,1]),
		\end{align*}
		and then set
		$x_k=\begin{cases}x'&\text{if } r\leq g(x')/g(x_{k-1})\\ x_{k-1}&\text{else}\end{cases}.$
	\item Iterate the previous step until the desired number of samples
		from the steady-state have been aquired.
\end{enumerate}
Intuitively this makes sense; we move from the previous 
location $x_{k-1}$ to the proposed location $x'$ with a 
probability that prefers a higher density of $f$, 
characterized by $f(x')/f(x_{k-1})=g(x')/g(x_{k-1})$.
Therefore samples will end up in the densest regions of 
$f(x)$.

The simulation method used in our paper is Hamiltonian Monte carlo (HMC) 
which is just a more sophisticated MCMC method~\cite{neal_mcmc_2012}.
Here, the term Hamiltonian is used in the classical context.
The unknown distribution $f(x)$ is treated as being the Boltzmann
distribution of some energy function over states in the sample space.
The proposal for the next step in the Markov chain simulation is 
drawn by simulating the dynamics of this Hamiltonian system using
the previous sample as the starting point and a random initial momentum
for some amount of time --- the endpoint of the trajectory is 
the proposal.
This results in very large steps and greatly decreases the mixing time;
a well tuned HMC sampler has nearly no correlation between adjacent 
points.
The main improvement made by the No-U-Turns sampler was to introduce an
automatic way to determine how long to simulate each Hamiltonian 
trajectory for~\cite{hoffman_no-u-turn_2014}.
A recent conceptual tutorial on Hamiltonian Monte Carlo has been provided by \citet{bet_conceptual_2017}.

\section{Beta Reparameterizations}
\label{app:dist-reparam}

In this section we provide some useful reparameterizations of the 
beta distribution, along with their inverses.
As noted in \autoref{eq:beta-pdf} of the main body,
a beta distribution $\betadist(\alpha,\beta)$ has a density function given by
\begin{equation}
    \pdf{\betadist}{q} = \frac{q^{\alpha-1}(1-q)^{\beta-1}}{\betafun{\alpha,\beta}}
\end{equation}
for any $q\in [0,1]$.
The normalization constant $\betafun{\alpha,\beta}$ 
is the beta function, which is defined in terms of the gamma function,
$\betafun{\alpha,\beta}=\frac{\Gamma(\alpha)\Gamma(\beta)}{\Gamma(\alpha+\beta)}$.
The parameters $\alpha$ and $\beta$ must both be positive.
Its mean and variance are given by 
\begin{subequations}
    \begin{align}
        \mu&=\frac{\alpha}{\alpha+\beta} \\
        \sigma^2&=\frac{\alpha\beta}{(\alpha+\beta)^2(\alpha+\beta+1)}
    \end{align}
\end{subequations}
respectively.
The conditions $\alpha,\beta>0$ are exactly equivalent to the conditions
$0<\mu<1$ and $0<\sigma^2<\mu(1-\mu)$.

The parameters $\alpha$ and $\beta$ have operational interpretations in 
terms of `prior observations'; $\alpha$ is the number of prior observations
of heads, and $\beta$ is the number of prior observations of tails.
Therefore, for example, the uniform prior $\betadist(1,1)$ asserts that
two prior observations have been made: one of heads, and one of tails.
not entirely intuitive.
In \autoref{tab:beta-reparams}, four reparameterizations along 
with their inverse transformations are given.

\begin{table*}[t]
  \centering
  \scriptsize
  \begin{tabularx}{\textwidth}{lllll}
    Parameters & Bounds & Transform & Inverse Transform & Variance \\
    \hline \\
    $(\mu, \sigma^2)$ 
        & $\mu\in(0,1)$ 
        & $\mu=\alpha/(\alpha+\beta)$ 
        & $\alpha = \mu^2(1-\mu)/\sigma^2 - \mu$ 
        & \\
        & $\sigma^2\in (0,\mu(1-\mu))$ ~~~
        & $\sigma^2=\alpha\beta/((\alpha+\beta)^2(\alpha+\beta+1))$~~~
        & $\beta = \mu(1-\mu)^2/\sigma^2 - (1-\mu)$ ~~~
        & $\sigma^2$ \\
    \hline \\
    $(\mu, \mu_2)$ 
        & $\mu\in(0,1)$ 
        & $\mu=\alpha/(\alpha+\beta)$ 
        & $\alpha = \mu(\mu-\mu_2)/(\mu_2-\mu^2)$ 
        & \\
        & $\mu_2\in (\mu^2,\mu)$ 
        & $\mu_2=\alpha(1+\alpha)/(\alpha+\beta)(1+\alpha+\beta)$
        & $\beta = (1-\mu)(\mu-\mu_2)/(\mu_2-\mu^2)$
        & $\sigma^2 = \mu_2 - \mu^2$ \\
    \hline \\
    $(\mu, t)$ 
        & $\mu\in(0,1)$ 
        & $\mu=\alpha/(\alpha+\beta)$ 
        & $\alpha = \mu(1/t-1)$ 
        & \\
        & $t\in (0,1)$ 
        & $t=1/(1+\alpha+\beta)$
        & $\beta = (1-\mu)(1-t)/t$ 
        & $\sigma^2=t \mu(1-\mu)$ \\
    \hline \\
    $(\mu, r)$ 
        & $\mu\in(0,1)$ 
        & $\mu=\alpha/(\alpha+\beta)$ 
        & $\alpha = 1/(r-r\mu)-\mu$ 
        & \\
        & $r\in (0,1)$ 
        & $r=(\alpha+\beta)^2/(\alpha\beta(1+\alpha+\beta))$
        & $\beta = 1/(r\mu)+\mu-1$ 
        & $\sigma^2=r \mu^2(1-\mu)^2$ \\
    \hline \\
    $(\mu, s)$ 
        & $\mu\in(0,1)$ 
        & $\mu=\alpha/(\alpha+\beta)$ 
        & $\alpha = s \mu$ 
        & \\
        & $s\in (0,\infty)$ 
        & $s=\alpha+\beta$
        & $\beta = s (1-\mu)$ 
        & $\sigma^2=\mu(1-\mu)/(s+1)$ \\
  \end{tabularx}
  \caption{Five reparameterizations of the beta distribution 
    $\betadist(\alpha,\beta)$. The first two, $(\mu,\sigma^2)$ and $(\mu,\mu_2)$, simply 
    reparameterizes into mean and variance (or second moment), which yields 
    non-rectangular bounds.
    The other three parameterizations have rectangular bounds.
    In the parameterization $(\mu,t)$, $t$ represents the fraction of the maximum possible
    variance given a mean $\mu$.
    Conversely, in the parameterization $(\mu,s)$, 
    we have $t=1/(s+1)$ so that large $s$ corresponds to small variance.
    The parameterization $(\mu,r)$ is the only one that does 
    not allow the full range of variance -- the maximum possible variance 
    (assuming $0<r<1$) is $\mu^2(1-\mu)^2$. This
    prevents build-up of mass at the boundaries $0$ and $1$ by forcing both 
    the constraints $\alpha>1$ and $\beta>1$.
  }
  \label{tab:beta-reparams}
\end{table*}

\section{Priors on Heavily Biased Coins}

In all of the examples in the main text we used an uninformative uniform prior
on probability parameters such as $A$, $B$, and $p$ in the standard
RB protocol.
There may be situations, especially in low data regimes, where
incorporating prior knowledge has a noticeable effect on the posterior
width, thereby reducing the necessary amount of data needed to 
attain a desired credibility lower bound.

For this purpose, we suggest a two-parameter family of distribution with 
support on $[0,1]$ which we call \textit{probably at least} (PAL).
A member of this family with parameters $0<p_0<1$ and $0<z<p_0$ has a 
continuous density function given by
\begin{align}
    \pdf{\PAL}{x}
        &= \frac{1-z}{1-p_0}
        \begin{cases}
            \left(\frac{x}{p_0}\right)^{\frac{p_0-z}{z(1-p_0)}} & x < p_0 \\
            1 & x \geq p_0
        \end{cases}.
\end{align}
This distribution is a sort of hedged version of $\Unif{[p_0,1]}$ that 
admits a finite but decreasing probability that $x < p_0$.
Indeed, it is parameterized so that the probability $0<x<p_0$ is equal 
to $z$.
Observe that $p_0=z$ gives $\Unif{[0,1]}$, and $z\rightarrow 0$ approaches
$\Unif{[p_0,1]}$.

If the discontinuity of the derivative of this prior at $x=p_0$
poses a problem 
for the sampler at hand, this distribution can be smoothed over as
follows:
\begin{align}
    g(x)
        &= \frac{p_0^2(2-z)+2xz+p_0(2x+z)}{zp_0(1-p_0)} \nonumber \\
    \pdf{\PAL'}{x}
        &= \frac{1-z}{1-p_0}
        \begin{cases}
            g(x)\left(\frac{x}{p_0}\right)^{2\frac{p_0-z}{z(1-p_0)}} 
            & x < p_0 \\
            1 & x \geq p_0
        \end{cases}.
\end{align}
The parameters $p_0$ and $z$ have the same interpretations, but now 
the decaying piece moves smoothly into the constant piece, at the cost
of a bit more complexity.

\section{A Reparameterization of LRB}
\label{app:reparam-leakage-rb}

Consider SPAM configurations $e=(\lambda,i)\in\Eset$ where $E_\lambda$ is 
a measurement operator, and $\rho_i$ is an initial state.
LRB as described in \cite{wood_quantification_2017} has a first moment
tying function $T(1,M,e,\vec{x}_T)$ defined by
\begin{align}    
        & \frac{1}{L_1+L_2}\Tr\left[
            E_\lambda^\dagger
            \E\left(
                L_2 \frac{\I_1}{d_1} + L_1 \frac{\I_2}{d_2}
            \right)
        \right] \nonumber \\
        &\quad + 
           \left(
            \frac{L_1}{L_1+L_2}-p_i
        \right)
        \Tr\left[
            E_\lambda^\dagger
            \E \left(
                \frac{\I_1}{d_1} - \frac{I_2}{d_2}
            \right)
        \right]
        \lambda_1^M \nonumber \\
        &\quad +
        (1-p_i)
        \Tr\left[
            E_\lambda^\dagger
            \E \left(
                \rho_i' - \frac{\I_1}{d_1}
            \right)
        \right]
        \lambda_2^M
\end{align}
where $\E$ is the gate independent noise acting 
on $\mathcal{X}_1\oplus\mathcal{X}_2$, with $\dim\mathcal{X}_k=d_k$
and $\I_k=\I_{\mathcal{X}_k}$ for $k=1,2$.
Here, $p_i:=\Tr[\I_1 \rho_i]$
and $\rho_i':=\I_1 \rho_i \I_1/(1-p_i)$.
The protocol recommends choosing $\lambda=0,...,d_1-1$ with
$E_\lambda\approx\ketbra{\lambda}$ and $i=0$ with $\rho_0\approx \ketbra{0}$.
Other quantities are defined as 
\begin{subequations}
    \begin{align}
        L_1 &= 1 - \Tr\E(\I_1/d_1)\I_1 \\
        L_2 &= \Tr\E(\I_2/d_2)\I_1 \\
        \lambda_1 &= 1-L_1-L_2 \\
        \lambda_2 &= \mu_1(1-L_1)=\frac{d_1 \overline{F}(\E)-(1-L_1)}{d_1-1}
    \end{align}
\end{subequations}
with $\overline{F}(\E)$ the average gate fidelity of $\E$ averaged 
over states in $\X_1$.
$L_1$, called the \textit{leakage}, measures $\E$'s average loss of 
population from $\mathcal{X}_1$ into $\mathcal{X}_2$, and $L_2$, called 
the \textit{seepage}, measures the reverse effect.
Some easy bounds on these parameters include
\begin{subequations}
    \begin{align}
        L_1,L_2 &\geq 0 \\
        L_1+L_2 &\leq 1.
    \end{align}
\end{subequations}

Wood and Gambetta suggest extracting the parameters of interest, 
$(L_1,L_2,\overline{F})$, 
as follows.
First, the data are summed over $\lambda$ and the sample mean is taken
over sequences $I$ and sequence repetitions $N$.
Under this sum the third term of the tying function,
$\sum_\lambda \Tr\left[
    E_\lambda^\dagger \E \left(
        \rho_0' - \frac{\I_1}{d_1}
    \right)
\right]\approx0$,
approximately cancels out leaving a single exponential term
of base $\lambda_1$.
Fitting to this curve yields $\lambda_1$ and hence $L_1+L_2$,
and combining this with the constant offset of the curve,
$\sum_\lambda \frac{1}{L_1+L_2}\Tr\left[
    E_\lambda^\dagger \E\left(
        L_2 \frac{\I_1}{d_1} + L_1 \frac{\I_2}{d_2}
\right)\right]
\approx \frac{L_2}{L_1+L_2}$,
we can separate to get $L_1$ and $L_2$.
Note that this protocol is not truly SPAM free because part of
the inference relies on the constant term which contains SPAM parameters.
Next we go back to the unsummed data, plug in our estimate of 
$\lambda_1$, and fit to $\lambda_2$ to deduce $\overline{F}$.

In our scheme, we are able to process the data all at once, instead 
of this two step fitting procedure.
It is helpful to rewrite the tying function a bit to make it 
a bit more clear what all of the independent parameters are.
We generalize the protocol to possibly use 
multiple initial states $\rho_i$,
$i=0,...,d-1$.
Then if we define $A_\lambda=\Tr[E_\lambda^\dagger\E(\I_1/d_1)]$,
$B_\lambda=\Tr[E_\lambda^\dagger\E(\I_2/d_2)]$, and
$C_{i,\lambda}=\Tr[E_\lambda^\dagger\E(\rho_i)]$ the tying function
is expressed as
\begin{align}
    T(1,M,(\lambda,i),\vec{x}_T) &=
    \frac{L_2 A_\lambda + L_1 B_\lambda}{L_1+L_2} \nonumber \\
    &\quad + \left(\frac{L_1}{L_1+L_2}-p_i\right)
        (A_\lambda-B_\lambda)\lambda_1^M \nonumber \\
    &\quad + (1-p_i)(C_{i,\lambda}-A_\lambda)\lambda_2^M.
\end{align}

There are two reasons that one might prefer to use an orthogonal
basis of pure initial states with one measurement operator, 
rather than vice versa, as suggested in the LRB paper.
The first is that it requires fewer nuisance parameters --- both 
$A_\lambda$ and $B_\lambda$ depend on the measurement but not 
the initial state.
The second is that the offset term 
$\frac{L_2 A_\lambda + L_1 B_\lambda}{L_1+L_2}$
is exactly equal for all experiments (under the assumption of gate-independent noise), which means it can effectively 
be measured independently by including very long sequence lengths 
in the data collection.

In \autoref{sec:lrb} of the main text, we used one measurement operator,
$M=0.99999\ketbra{0}$, and two initial states,
$\rho_0=0.9999\ketbra{0}$ and $\rho_1=0.9995\ketbra{1}$.
The two prior distributions used for tying parameters were
\begin{subequations}
\begin{align}
    L_1,L_2,\cdot &\sim \dirichlet(1,1,100) \\
    \mu_1, A_\lambda, B_\lambda, C_{i,\lambda} &\sim \uniform([0,1]).
\end{align}
\end{subequations}
and
\begin{subequations}
\begin{align}
    L_1,L_2,\cdot &\sim \dirichlet(1,1,100) \\
    \mu_1,  C_{i,\lambda} &\sim \uniform([0,1]) \\
    A_\lambda &\sim \betadist(100,100) \\
    B_\lambda &\sim \betadist(1,100)
\end{align}
\end{subequations}
labeled `Flat SPAM prior' and `Tighter SPAM prior' in \autoref{fig:lrb},
respectively.
The Dirichlet distribution on $L_1$ and $L_2$ was chosen because 
of the additive constraint $L_1+L_2\leq1$ means that 
the triple $(L_1,L_2,1-L_1-L_2)$ is a probability vector.
The variable $0\leq \mu_1 \leq 1$ has the interpretation of the depolarizing 
parameter of $\E$ restricted to $\X_1$.

\end{document}